# FACIAL STRUCTURES FOR VARIOUS NOTIONS OF POSITIVITY AND APPLICATIONS TO THE THEORY OF ENTANGLEMENT

SEUNG-HYEOK KYE

DEPARTMENT OF MATHEMATICS
SEOUL NATIONAL UNIVERSITY
SEOUL 151-742, KOREA

ABSTRACT. In this expository note, we explain facial structures for the convex cones consisting of positive linear maps, completely positive linear maps, decomposable positive linear maps between matrix algebras, respectively. These will be applied to study the notions of entangled edge states with positive partial transposes and optimality of entanglement witnesses.

The notion of quantum entanglement has been one of the key research areas of quantum physics since the nineties, in relation with possible applications to quantum information and quantum computation theory. Since the set of all separable states is a convex set, convex geometry may be one of the mathematical framework to study these notions. In fact, the convex duality between various cones in tensor product spaces and linear mapping spaces are very useful to characterize the various notions for entanglement, and has been used implicitly by physicists to detect entanglement.

One of the best way to understand the whole structures of a given convex set is to characterize the lattice of all faces. The duality plays a key role for this purpose, since it gives us a simple way to describe exposed faces among all faces. It turns out that important notions like separability, Schmidt numbers and positive partial transpose may be explained as the dual objects of various notions of positivity. So, we begin this note to introduce several notions of positivity of linear maps between matrix algebras including $s$-positivity, complete positivity, complete copositivity and decomposability. We will introduce the several notions of entanglement in terms of dual objects of these notions.

It is easy to characterize the facial structures for complete positivity with which it is also possible to describe faces for decomposable positive maps. It is also possible to determine the boundary structures for positive linear maps, although it is very difficult to know the whole facial structures for those.

One of the main theme in the theory of entanglement is to determine if a given state is separable or not. Since every separable state is of positive partial transpose, it is important to understand the facial structures for PPT states. In this context, the notion of PPT

1991 *Mathematics Subject Classification.* 81P15, 15A30, 46L05.
partially supported by NRFK 2011-0001250.

entangled edge states plays an important role. In Section 7, We construct various types of edge states and classify them in low dimensional cases.

Another topic of this note is the notion of optimal entanglement witnesses, which will be also explained in Section 8 in terms of facial structures for positive maps. Especially, the notion of spanning property will be explained in terms of faces, which seems to be new. See Proposition 8.3. We also exhibit examples to distinguish several notions of optimality, and discuss the role of exposed positive maps as entanglement witnesses.

This note touches very small parts of the whole aspects of the theory of entanglement, which attracts recently many mathematicians including functional analysts. See [65], [66], [94], [96] for approaches using the theory of operator systems and operator spaces, and see [5], [100] for measure theoretic approach, for examples.

This note is an outcome of the series of author's lectures given at Ritsumeikan University in October, 2011. He is very grateful to Professor Hiroyuki Osaka for his warm hospitality during his stay there as well as stimulating discussion on the topics. He is also grateful to all audience, especially to Professor Jun Tomiyama whose comments were very useful to prepare this note. Special thanks are due to Professors Kil-Chan Ha and Kyung Hoon Han for their various useful comments on the draft. Finally, the author appreciate the referee's useful suggestions and careful reading.

## Contents

## 1. Various notions of positivity

A linear map $\phi : A \to B$ between $C^*$-algebras $A$ and $B$ is said to be *positive* if it sends the convex cone $A^+$ of all positive elements into the cone $B^+$. We denote by $M_s(A)$ the $C^*$-algebra of all $s \times s$ matrices over $A$. If the linear map

(1) $$M_s(A) \to M_s(B) : [x_{ij}]_{i,j=1}^s \mapsto [\phi(x_{ij})]_{i,j=1}^s$$

is positive then we say that $\phi$ is *s-positive*. Throughout this note, we use the tensor notation with which $M_s(A)$ will be the tensor product $M_s \otimes A$ of the $C^*$-algebra $M_s$ of all $s \times s$ matrices over the complex field and the $C^*$-algebra $A$. Then the block matrix $[x_{ij}]_{i,j=1}^s \in M_s(A)$ corresponds to $\sum_{i,j=1}^s e_{ij} \otimes x_{ij} \in M_s \otimes A$, where $\{e_{ij}\}$ denotes the usual matrix units. With this notation, the map (1) can be written by

$$\mathrm{id}_s \otimes \phi : M_s \otimes A \to M_s \otimes B : \sum e_{ij} \otimes x_{ij} \mapsto \sum e_{ij} \otimes \phi(x_{ij}),$$

where $\mathrm{id}_s$ denotes the identity map of the $C^*$-algebra $M_s$. We denote by $\mathbb{P}_s[A, B]$ the convex cone of all $s$-positive linear maps from $A$ into $B$. If $\phi$ is $s$-positive for each $s = 1, 2, \ldots$, then we say that $\phi$ is *completely positive*.

The transpose map

$$\mathrm{tp}_s : M_s \to M_s : x \mapsto x^{\mathrm{t}}$$

is a typical example of a positive linear map which is not completely positive. We look at the case of $s = 2$. We see that the map

$$\mathrm{id}_2 \otimes \mathrm{tp}_2 : M_2 \otimes M_2 \to M_2 \otimes M_2$$

is not positive. Indeed, it send the positive semi-definite matrix

(2) $$\sum_{i,j=1}^2 e_{ij} \otimes e_{ij} = \begin{pmatrix} 1 & 0 & 0 & 1 \\ 0 & 0 & 0 & 0 \\ 0 & 0 & 0 & 0 \\ 1 & 0 & 0 & 1 \end{pmatrix}$$

of $M_2 \otimes M_2$ to the matrix

$$\sum_{i,j=1}^2 e_{ij} \otimes \mathrm{tp}_2(e_{ij}) = \sum_{i,j=1}^2 e_{ij} \otimes e_{ji} = \begin{pmatrix} 1 & 0 & 0 & 0 \\ 0 & 0 & 1 & 0 \\ 0 & 1 & 0 & 0 \\ 0 & 0 & 0 & 1 \end{pmatrix}$$

which is not positive semi-definite. So, we see that the transpose map is not 2-positive.

A linear map $\phi : A \to B$ is said to be *s-copositive* if the map

$$\mathrm{tp}_s \otimes \phi : M_s \otimes A \to M_s \otimes B$$

is positive. The convex cone of all $s$-positive maps from $A$ into $B$ will be denoted by $\mathbb{P}^s[A, B]$. If $\phi$ is $s$-copositive for each $s = 1, 2, \ldots$, then we say that $\phi$ is *completely copositive*. A positive linear map is said to be *decomposable* if it is the sum of a completely positive map and a completely copositive map.

For a given $m \times n$ matrix $V$, the map $\phi_V : M_m \to M_n$ defined by

$$\phi_V : X \mapsto V^*XV, \qquad X \in M_m$$

is a typical example of a completely positive linear map. Indeed, we have

$$(\mathrm{id}_s \otimes \phi_V)(Y \otimes X) = Y \otimes (V^*XV) = (I_s \otimes V)^*(Y \otimes X)(I_s \otimes V)$$
3

for every $Y \otimes X \in M_s \otimes M_m$, where $I_s$ denotes the identity matrix of $M_s$. On the other hand, the map defined by

$$\phi^V : X \mapsto V^* X^{\mathrm{t}} V, \qquad X \in M_m$$

is a completely copositive map. For a finite family $\mathcal{V}$ of $m \times n$ matrices, the map

$$\phi_{\mathcal{V}} : X \mapsto \sum_{V \in \mathcal{V}} V^* X V$$

is also a completely positive map. Actually, the following theorem [25], [74] tells us that they exhaust all completely positive linear maps between matrix algebras.

**Theorem 1.1.** *For a linear map $\phi : M_m \to M_n$, the following are equivalent:*

(i) *$\phi$ is completely positive.*
(ii) *$\phi$ is $m \wedge n$-positive, where $m \wedge n$ denotes the minimum of $m$ and $n$.*
(iii) *The matrix*

$$C_\phi := \sum_{i,j=1}^m e_{ij} \otimes \phi(e_{ij}) \in M_m \otimes M_n$$

*is positive semi-definite.*
(iv) *There exists a linearly independent family $\mathcal{V}$ of $m \times n$ matrices such that $\phi = \phi_{\mathcal{V}}$.*

We call $C_\phi \in M_m \otimes M_n$ the *Choi matrix* of the linear map $\phi$ from $M_m$ into $M_n$. The correspondence $\phi \mapsto C_\phi$ from the space $\mathcal{L}(M_m, M_n)$ of all linear maps onto the space $M_m \otimes M_n$ is called the *Jamiołkowski-Choi isomorphism*. See [64].

For an $m \times n$ matrix $V$, we denote by $V_i$ the $i$th row. Then we have

$$V^* e_{ij} V = V_i^* V_j \in M_n,$$

for the matrix units $\{e_{ij} : i, j = 1, \ldots, m\}$ of $M_m$. Therefore, the Choi matrix $C_{\phi_V}$ of the map $\phi_V$ is given by

$$(3) \qquad C_{\phi_V} = \sum_{i,j=1}^m e_{ij} \otimes V_i^* V_j = \left(\sum_{i=1}^m e_i \otimes V_i^*\right)\left(\sum_{j=1}^m e_j \otimes V_j^*\right)^* \in M_m \otimes M_n.$$

This is the rank one projector onto the vector $\sum_{i=1}^m e_i \otimes V_i^* \in \mathbb{C}^m \otimes \mathbb{C}^n$, where $\{e_i\}$ denotes the usual orthogonal basis. This actually proves Theorem 1.1. Indeed, the $m$-positivity of $\phi$ implies that the matrix

$$C_\phi = (\mathrm{id}_m \otimes \phi) \left(\sum_{i,j=1}^m e_{ij} \otimes e_{ij}\right)$$

is positive semi-definite, since the matrix

$$\sum_{i,j=1}^m e_{ij} \otimes e_{ij} = \left(\sum_{i=1}^m e_i \otimes e_i\right)\left(\sum_{j=1}^m e_j \otimes e_j\right)^* \in M_m \otimes M_m$$

is positive semi-definite. If $C_\phi$ is positive semi-definite then we may write $C_\phi = \sum_\iota z_\iota z_\iota^*$ with $z_\iota \in \mathbb{C}^m \otimes \mathbb{C}^n$. This gives us the expression $\phi = \sum_\iota \phi_{V_\iota}$ by (3). Finally, it is easy to that $\phi : M_m \to M_n$ is $m$-positive if and only if it is $n$-positive, considering the dual map from $M_n$ into $M_m$.

Many efforts had been made to find examples which may distinguish various notions of positivity. For nonnegative real numbers $a, b$ and $c$, we consider the linear map

$$\Phi[a,b,c] : M_3 \to M_3$$

defined by

$$(4) \quad \Phi[a,b,c](X) = \begin{pmatrix} ax_{11} + bx_{22} + cx_{33} & -x_{12} & -x_{13} \\ -x_{21} & cx_{11} + ax_{22} + bx_{33} & -x_{23} \\ -x_{31} & -x_{32} & bx_{11} + cx_{22} + ax_{33} \end{pmatrix}$$

for $X = [x_{ij}] \in M_3$, as was introduced in [21]. The first example of a map of this type was given by Choi [24], who showed that the map $\Phi[1,2,2]$ is a 2-positive linear map which is not completely positive. This is the first example to distinguish $n$-positivities for different $n$'s. See also [117, 123]. The map $\Phi[1,0,\mu]$ with $\mu \geq 1$ is also the first example of an indecomposable positive linear map given by Choi [26]. The map $\Phi[1,0,1]$, which is usually called the Choi map, was shown [28] to generate an extremal ray of the cone $\mathbb{P}_1$. Furthermore, it turns out [118] that this map $\Phi[1,0,1]$ is an atom, that is, it is not the sum of a 2-positive map and a 2-copositive map. See also [41]. We summarize the results in [21] as follows:

**Theorem 1.2.** *Let $a, b$ and $c$ be nonnegative real numbers. Then the map $\Phi[a,b,c]$ is*

  (i) *positive if and only if $a + b + c \geq 2$ and $0 \leq a \leq 1 \to bc \geq (1-a)^2$,*
  (ii) *2-positive if and only if $a \geq 2$ or $[1 \leq a < 2] \wedge [bc \geq (2-a)(b+c)]$,*
  (iii) *completely positive if and only if $a \geq 2$,*
  (iv) *2-copositive if and only if completely copositive if and only if $bc \geq 1$,*
  (v) *decomposable if and only if $0 \leq a \leq 2 \to bc \geq \left(\frac{2-a}{2}\right)^2$.*

Here, $p \to q$ means that $q$ holds in case of $p$. See Figure 1. We note that the Choi matrix of the map $\Phi[a,b,c]$ is given by

$$(5) \quad A[a,b,c] := \begin{pmatrix} a & \cdot & \cdot & \cdot & -1 & \cdot & \cdot & \cdot & -1 \\ \cdot & c & \cdot & \cdot & \cdot & \cdot & \cdot & \cdot & \cdot \\ \cdot & \cdot & b & \cdot & \cdot & \cdot & \cdot & \cdot & \cdot \\ \cdot & \cdot & \cdot & b & \cdot & \cdot & \cdot & \cdot & \cdot \\ -1 & \cdot & \cdot & \cdot & a & \cdot & \cdot & \cdot & -1 \\ \cdot & \cdot & \cdot & \cdot & \cdot & c & \cdot & \cdot & \cdot \\ \cdot & \cdot & \cdot & \cdot & \cdot & \cdot & c & \cdot & \cdot \\ \cdot & \cdot & \cdot & \cdot & \cdot & \cdot & \cdot & b & \cdot \\ -1 & \cdot & \cdot & \cdot & -1 & \cdot & \cdot & \cdot & a \end{pmatrix}.$$

Note that $\Phi[a,b,c]$ is completely positive if and only if $A[a,b,c]$ is positive semi-definite if and only if $a \geq 2$. For example, the map $\Phi[2,0,0]$ can be written by

$$\Phi[2,0,0] = \phi_{e_{11}-e_{22}} + \phi_{e_{22}-e_{33}} + \phi_{e_{33}-e_{11}}.$$

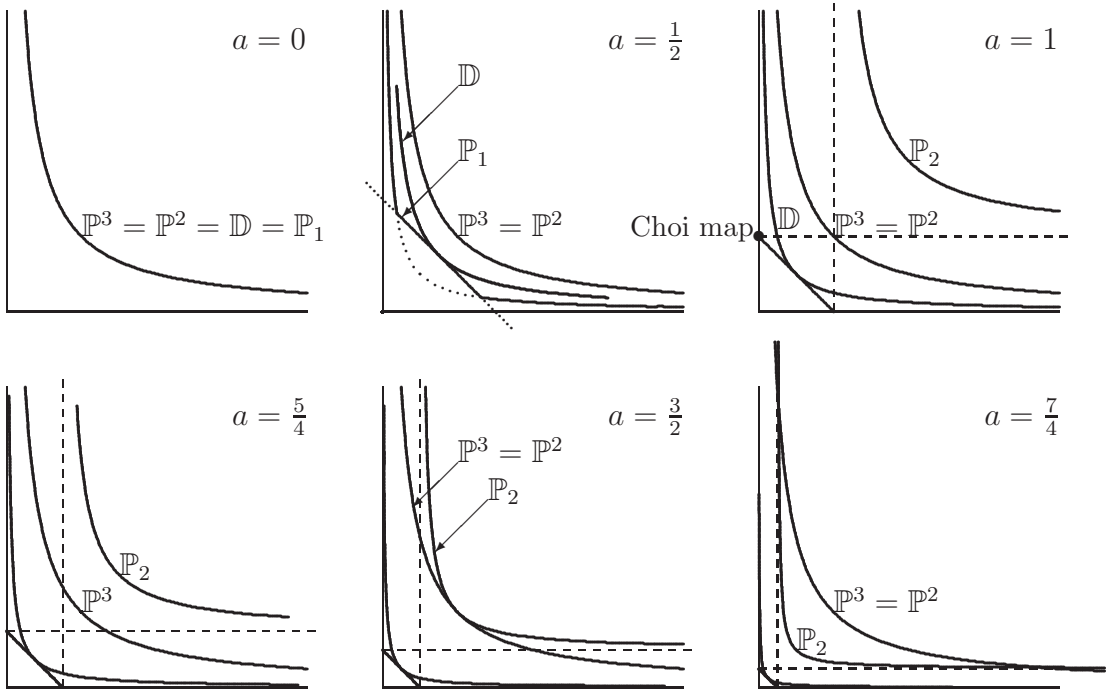

FIGURE 1. The horizontal and vertical axes represent $b$ and $c$ axes, respectively. The hyperbola denoting $\mathbb{P}_3$ is given by the equation $bc = 1$ in each picture, and the Choi map is represented by the point $(b, c) = (0, 1)$. When $a = \frac{3}{2}$ the asymptotic lines for the hyperbola denoting $\mathbb{P}_2$ are $b = \frac{1}{2}$ and $c = \frac{1}{2}$.

On the other hand, the completely copositive map $\Phi[0, 1, 1]$ may be written by

$$\Phi[0, 1, 1] = \phi^{e_{12}-e_{21}} + \phi^{e_{23}-e_{32}} + \phi^{e_{31}-e_{13}}.$$

We note that there are another variants of the Choi map as was considered in [75]. Some of them, parameterized by three real variables, were shown [92] to generate extreme rays. See also [3], [14], [32], [33], [34], [51], [91], [105], [118], [130] for another variations of the Choi map. One may consider positive maps which fix diagonals. It turns out [76] that every positive map between $M_3$ fixing diagonals becomes decomposable. But it is known [70] that there exist a diagonal fixing positive maps between $M_4$ which is not decomposable.

It was shown by Woronowicz [127] that every positive linear map from $M_2$ into $M_n$ is decomposable if and only if $n \leq 3$. The first explicit example of indecomposable positive linear map between $M_2$ and $M_4$ was given in [128]. See also [119]. We refer to [97] and [98] for examples of indecomposable positive linear maps between $M_4$. For more extensive examples of indecomposable positive linear maps, we refer to [31].

## 2. Duality

Let $X$ and $Y$ be finite-dimensional normed real spaces, which are dual to each other with respect to a bilinear pairing $\langle \, , \, \rangle$. For a subset $C$ of $X$, we define the *dual cone* $C^\circ$ by

$$C^\circ = \{y \in Y : \langle x, y \rangle \geq 0 \text{ for each } x \in C\},$$

and the dual cone $D^\circ \subset X$ similarly for a subset $D$ of $Y$. It is clear that $C^{\circ\circ}$ is the closed convex cone generated by $C$. Therefore, every closed convex cone $C$ of $X$ is the dual cone of $C^\circ \subset Y$, and it is determined by the intersection of 'half-spaces' $\{x : \langle x, y \rangle \geq 0\}$ induced by elements $y$ in $C^\circ$.

We denote by $\mathcal{B}(\mathcal{H})$ and $\mathcal{T}(\mathcal{H})$ the space of all bounded linear operators and trace class operators on a Hilbert space $\mathcal{H}$, respectively. We use the duality between the space $\mathcal{B}(A, \mathcal{B}(\mathcal{H}))$ of all bounded linear operators from a $C^*$-algebra $A$ into $\mathcal{B}(\mathcal{H})$ and the projective tensor product $A \hat{\otimes} \mathcal{T}(\mathcal{H})$ given by

$$\langle x \otimes y, \phi \rangle = \operatorname{Tr}(\phi(x)y^{\mathrm{t}}), \qquad x \in A, \ y \in \mathcal{T}(\mathcal{H}), \ \phi \in \mathcal{B}(A, \mathcal{B}(\mathcal{H})),$$

where Tr denotes the usual trace. This duality was used by Woronowicz [127] to show that every positive linear map from the matrix algebra $M_2$ into $M_n$ is decomposable if and only if $n \leq 3$. The above duality is also useful to study extendibility of positive linear maps as was considered by Størmer [113]. The predual cones of $\mathbb{P}_s[A, \mathcal{B}(\mathcal{H})]$ and $\mathbb{P}^s[A, \mathcal{B}(\mathcal{H})]$ with respect to the above pairing have been determined by Itoh [62].

If we restrict ourselves to the cases of matrix algebras, this gives rise to the duality between the space $M_m \otimes M_n$ and the space $\mathcal{L}(M_m, M_n)$. For $A = \sum_{i,j=1}^m e_{ij} \otimes a_{ij} \in M_m \otimes M_n$ and a linear map $\phi \in \mathcal{L}(M_m, M_n)$, we have

$$\langle A, \phi \rangle = \sum_{i,j=1}^m \operatorname{Tr}(\phi(e_{ij}) a_{ij}^{\mathrm{t}}) = \sum_{i,j=1}^m \langle a_{ij}, \phi(e_{ij}) \rangle,$$

where the bilinear form in the right-side is given by $\langle X, Y \rangle = \operatorname{Tr}(YX^{\mathrm{t}})$ for $X, Y \in M_n$. Therefore, this pairing is nothing but

$$\langle A, \phi \rangle = \operatorname{Tr}(AC_\phi^{\mathrm{t}}) = \operatorname{Tr}(C_\phi A^{\mathrm{t}})$$

for two matrices $A$ and $C_\phi$ in $M_m \otimes M_n$ with the usual trace.

Now, we proceed to determine the dual cone of the cone $\mathbb{P}_s[M_m, M_n]$. Every vector $z \in \mathbb{C}^m \otimes \mathbb{C}^n$ may be written in a unique way as $z = \sum_{i=1}^m e_i \otimes z_i$ with $z_i \in \mathbb{C}^n$ for $i = 1, 2, \ldots, m$. We say that $z$ is an *s-simple vector* in $\mathbb{C}^m \otimes \mathbb{C}^n$ if the linear span of $\{z_1, \ldots, z_m\}$ has the dimension $\leq s$. A 1-simple vector is called a *product vector*.

For an $s$-simple vector $z = \sum_{i=1}^m e_i \otimes z_i \in \mathbb{C}^m \otimes \mathbb{C}^n$, take a generator $\{u_1, u_2, \ldots, u_s\}$ of the linear span of $\{z_1, z_2, \ldots, z_m\}$ in $\mathbb{C}^n$, and define $a_{ik} \in \mathbb{C}$, $a_k \in \mathbb{C}^m$ by

(6)
$$z_i = \sum_{k=1}^s a_{ik} u_k \in \mathbb{C}^n, \qquad i = 1, 2, \ldots, m,$$
$$a_k = \sum_{i=1}^m a_{ik} e_i \in \mathbb{C}^m, \qquad k = 1, 2, \ldots, s.$$

Then we have
$$zz^* = \sum_{i,j=1}^m e_{ij} \otimes z_i z_j^* \in M_m \otimes M_n, \qquad z_i z_j^* = \sum_{k,\ell=1}^s a_{ik} \bar{a}_{j\ell} u_k u_\ell^* \in M_n,$$

and so it follows that
$$\langle zz^*, \phi \rangle = \sum_{i,j=1}^m \langle z_i z_j^*, \phi(e_{ij}) \rangle$$
$$= \sum_{i,j=1}^m \sum_{k,\ell=1}^s a_{ik} \bar{a}_{j\ell} \langle u_k u_\ell^*, \phi(e_{ij}) \rangle = \sum_{i,j=1}^m \sum_{k,\ell=1}^s a_{ik} \bar{a}_{j\ell} (\phi(e_{ij}) \bar{u}_\ell | \bar{u}_k)_{\mathbb{C}^n}$$

where $(\ |\ )_{\mathbb{C}^n}$ denotes the inner product of $\mathbb{C}^n$ which is linear in the first variable and conjugate-linear in the second variable. Therefore, we have
$$\langle zz^*, \phi \rangle = \sum_{i,j=1}^m \sum_{k,\ell=1}^s a_{ik} \bar{a}_{j\ell} (e_{k\ell} \otimes \phi(e_{ij})) u | u)_{\mathbb{C}^s \otimes \mathbb{C}^n},$$

where

(7)
$$u = \sum_{k=1}^s e_k \otimes \bar{u}_k \in \mathbb{C}^s \otimes \mathbb{C}^n.$$

If we put

(8)
$$w = \sum_{k=1}^s e_k \otimes a_k \in \mathbb{C}^s \otimes \mathbb{C}^m,$$

then we have
$$(\mathrm{id}_s \otimes \phi)(ww^*) = \sum_{k,\ell=1}^s e_{k\ell} \otimes \phi(a_k a_\ell^*) = \sum_{k,\ell=1}^s \sum_{i,j=1}^m a_{ik} \bar{a}_{j\ell} e_{k\ell} \otimes \phi(e_{ij}).$$

Therefore, it follows that

(9)
$$\langle zz^*, \phi \rangle = ((\mathrm{id}_s \otimes \phi)(ww^*) u | u)_{\mathbb{C}^s \otimes \mathbb{C}^n}.$$

Assume that $\phi$ is $s$-positive and take an $s$-simple vector $z = \sum_{i=1}^m e_i \otimes z_i \in \mathbb{C}^m \otimes \mathbb{C}^n$. Then the identity (9) shows that $\langle zz^*, \phi \rangle \geq 0$. For the converse, assume that $\langle zz^*, \phi \rangle \geq 0$ for each $s$-simple vector $z \in \mathbb{C}^m \otimes \mathbb{C}^n$. For each $w \in \mathbb{C}^s \otimes \mathbb{C}^m$ and $u \in \mathbb{C}^s \otimes \mathbb{C}^n$ as in (7) and (8), we take $z_i \in \mathbb{C}^n$ as in the relations (6). Then, we see that $(\mathrm{id}_s \otimes \phi)(ww^*)$ is positive semi-definite by (9), and so $\mathrm{id}_s \otimes \phi$ is a positive linear map. In short, the map $\phi$ is $s$-positive if and only if $\langle zz^*, \phi \rangle \geq 0$ for each $s$-simple vector $z \in \mathbb{C}^m \otimes \mathbb{C}^n$.

For a matrix $x \otimes y \in M_m \otimes M_n$, the *partial transpose* is defined by
$$(x \otimes y)^\tau = x^{\mathrm{t}} \otimes y.$$
For a matrix $A = \sum_{i,j=1}^m e_{ij} \otimes x_{ij} \in M_m \otimes M_n$, the partial transpose $A^\tau$ of $A$ is given by
$$A^\tau = \sum_{i,j=1}^m e_{ij}^{\mathrm{t}} \otimes x_{ij} = \sum_{i,j=1}^m e_{ji} \otimes x_{ij} = \sum_{i,j=1}^m e_{ij} \otimes x_{ji}.$$
Therefore, the partial transpose is nothing but the block-wise transpose of the corresponding block matrix in $M_m(M_n)$. The same calculation shows the identity
$$\langle (zz^*)^\tau, \phi \rangle = ((\mathrm{tp}_s \otimes \phi)(\bar w \bar w^*) u | u)_{\mathbb{C}^s \otimes \mathbb{C}^n}$$
also holds. We summarize in the following [38]:

**Theorem 2.1.** *For a linear map $\phi : M_m \to M_n$, we have the following:*

(i) *The map $\phi$ is s-positive if and only if $\langle zz^*, \phi \rangle \geq 0$ for each s-simple vector $z \in \mathbb{C}^m \otimes \mathbb{C}^n$.*

(ii) *The map $\phi$ is s-copositive if and only if $\langle (zz^*)^\tau, \phi \rangle \geq 0$ for each s-simple vector $z \in \mathbb{C}^m \otimes \mathbb{C}^n$.*

For $s = 1, 2, \ldots, m \wedge n$, we define the convex cones $\mathbb{V}_s$ and $\mathbb{V}^s$ in $M_m \otimes M_n$ by
$$\mathbb{V}_s(M_m \otimes M_n) = \{zz^* : z \text{ is an } s\text{-simple vector in } \mathbb{C}^m \otimes \mathbb{C}^n\}^{\circ\circ},$$
$$\mathbb{V}^s(M_m \otimes M_n) = \{(zz^*)^\tau : z \text{ is an } s\text{-simple vector in } \mathbb{C}^m \otimes \mathbb{C}^n\}^{\circ\circ}.$$
Then Theorem 2.1 says that $(\mathbb{V}_s, \mathbb{P}_s)$ is a dual pair in the following sense:
$$\phi \in \mathbb{P}_s \iff \langle A, \phi \rangle \geq 0 \text{ for each } A \in \mathbb{V}_s,$$
$$A \in \mathbb{V}_s \iff \langle A, \phi \rangle \geq 0 \text{ for each } \phi \in \mathbb{P}_s,$$
and similarly for the pair $(\mathbb{V}^t, \mathbb{P}^t)$. We note that $\mathbb{V}_{m \wedge n}(M_m \otimes M_n)$ is nothing but the cone $(M_m \otimes M_n)^+$ of all positive semi-definite matrices in $M_m \otimes M_n$. We also note that the cone $\mathbb{P}_{m \wedge n}$ also corresponds to the cone $(M_m \otimes M_n)^+$ via the Jamiołkowski-Choi isomorphism by Theorem 1.1. Therefore, the duality between $\mathbb{V}_{m \wedge n}$ and $\mathbb{P}_{m \wedge n}$ is a restatement of the well-known fact that a matrix $A = [a_{ij}] \in M_N$ is positive semi-definite if and only if $\mathrm{Tr}\,(BA^{\mathrm{t}}) = \sum_{i,j=1}^N a_{ij} b_{ij} \geq 0$ for every positive semi-definite $B = [b_{ij}] \in M_N$. Dualities between cones may be explained by the following diagram together with inclusion relations between the cones:

(10)
$$\begin{array}{ccccccc}
\mathbb{V}_1 & \subset & \mathbb{V}_2 & \subset & \cdots & \subset & \mathbb{V}_{m \wedge n} = (M_m \otimes M_n)^+ \\
\updownarrow & & \updownarrow & & & & \updownarrow \\
\mathbb{P}_1 & \supset & \mathbb{P}_2 & \supset & \cdots & \supset & \mathbb{P}_{m \wedge n} \cong (M_m \otimes M_n)^+
\end{array}$$

where $\cong$ denotes the Jamiołkowski-Choi isomorphism. A linear map $\phi$ is said to be *super-positive* [4] or an *entanglement breaking channel* [56], [67] in the literature if $C_\phi$ belongs to the cone $\mathbb{V}_1$. On the other hand, a block matrix is said to be *block-positive* if it is the Choi matrix $C_\phi$ of a $\phi \in \mathbb{P}_1$. For more systematic approach to the duality together with the Jamiołkowski-Choi isomorphism, we refer to [109] and [132]. See also [106], [114], [115], and [116].

It is easy to see that
$$(C_1 + C_2)^\circ = C_1^\circ \cap C_2^\circ, \qquad (C_1 \cap C_2)^\circ = C_1^\circ + C_2^\circ,$$
whenever $C_1$ and $C_2$ are closed convex cones of $X$. Therefore, the following
$$(\mathbb{V}_s \cap \mathbb{V}^t, \mathbb{P}_s + \mathbb{P}^t)$$
is also a dual pair. Note that the cone $\mathbb{P}_{m \wedge n} + \mathbb{P}^{m \wedge n}$ consists of all decomposable maps, which will be denoted by $\mathbb{D}$:
$$\mathbb{D} := \mathbb{P}_{m \wedge n} + \mathbb{P}^{m \wedge n}.$$
Its dual cone $\mathbb{V}_{m \wedge n} \cap \mathbb{V}^{m \wedge n}$ will be denoted by $\mathbb{T}$:
$$\mathbb{T} := \{A \in (M_m \otimes M_n)^+ : A^\tau \in (M_m \otimes M_n)^+\}.$$
Then, we also have
$$\phi \in \mathbb{D} \iff \langle A, \phi \rangle \geq 0 \text{ for each } A \in \mathbb{T},$$
$$A \in \mathbb{T} \iff \langle A, \phi \rangle \geq 0 \text{ for each } \phi \in \mathbb{D}.$$
Elements in the cone $\mathbb{T}$ are said to be of *positive partial transpose* or *PPT*, in short. Note that the Choi matrix $C_\phi$ of a map $\phi \in \mathbb{P}_{m \wedge n} \cap \mathbb{P}^{m \wedge n}$ belongs to $\mathbb{T}$. Conversely, every element of $\mathbb{T}$ gives rise to a map which is both completely positive and completely copositive through the Jamiołkowski-Choi isomorphism. For example, the matrix $A[a, b, c]$ in (5) is of PPT if and only if $a \geq 2$ and $bc \geq 1$.

We also have the following diagram:

(11)
$$\begin{array}{ccccccc} \mathbb{V}_1 & \subset & \mathbb{T} & \subset & \mathbb{V}_{m \wedge n} & = & (M_m \otimes M_n)^+ \\ \updownarrow & & \updownarrow & & \updownarrow & & \\ \mathbb{P}_1 & \supset & \mathbb{D} & \supset & \mathbb{P}_{m \wedge n} & \cong & (M_m \otimes M_n)^+ \end{array}$$

We cannot combine (10) and (11) to draw a single diagram, since we do not know the inclusion relation between $\mathbb{T}$ and $\mathbb{V}_s$ when $1 < s < m \wedge n$. It was conjectured that
$$\mathbb{T}[M_3, M_3] \subset \mathbb{V}_2[M_3, M_3]$$
in [102]. Note that this is equivalent to claim the following relation
$$\mathbb{D}[M_3, M_3] \supset \mathbb{P}_2[M_3, M_3],$$
which is true for the maps $\Phi[a, b, c]$ by Theorem 1.2.

As another application of Theorem 2.1, we also have

(12) $$\phi \in \mathbb{P}^s[M_m, M_n] \iff \phi \circ \mathrm{tp}_m \in \mathbb{P}_s[M_m, M_n].$$

Indeed, from the identity
$$\langle x^t \otimes y, \phi \rangle = \mathrm{Tr}\,(\phi(x^t)y^t) = \mathrm{Tr}\,(((\phi \circ \mathrm{tp})(x))y^t) = \langle x \otimes y, \phi \circ \mathrm{tp} \rangle,$$

we have the following relation

(13) $$\langle A^\tau, \phi \rangle = \langle A, \phi \circ \mathrm{tp} \rangle,$$

from which the relation (12) follows. Note that we also have the relation

(14) $$\phi \in \mathbb{P}^s[M_m, M_n] \iff \mathrm{tp}_n \circ \phi \in \mathbb{P}_s[M_m, M_n]$$

by definition. Indeed, we have
$$\begin{aligned}
\mathrm{id}_s \otimes (\mathrm{tp}_n \circ \phi) &= (\mathrm{id}_s \otimes \mathrm{tp}_n) \circ (\mathrm{id}_s \otimes \phi) \\
&= (\mathrm{tp}_s \otimes \mathrm{tp}_n) \circ (\mathrm{tp}_s \otimes \mathrm{id}_n) \circ (\mathrm{id}_s \otimes \phi) = (\mathrm{tp}_s \otimes \mathrm{tp}_n) \circ (\mathrm{tp}_s \otimes \phi),
\end{aligned}$$
and $(\mathrm{tp}_s \otimes \mathrm{tp}_n)$ is the usual transpose map on $M_s \otimes M_n$.

## 3. Entanglement

Note that every density matrix $A$ in $M_\nu$ gives rise to a state of the $C^*$-algebra $M_\nu$ through $B \mapsto \mathrm{Tr}(AB^\mathrm{t})$. Therefore, every element of the cone $\mathbb{V}_{m \wedge n} = (M_m \otimes M_n)^+$ gives rise to a state of the $C^*$-algebra $M_m \otimes M_n$ if it is normalized. We say that a state in $\mathbb{V}_{m \wedge n}$ is said to be *separable* if it belongs to the smaller cone $\mathbb{V}_1$. Throughout this note, we ignore the normalization and call an element in the cone $\mathbb{V}_1$ to be separable. Therefore, a positive semi-definite matrix in $M_m \otimes M_n$ is separable if and only if it is a linear combination with positive coefficients of rank one projectors onto product vectors in $\mathbb{C}^m \otimes \mathbb{C}^n$. For a product vector $\xi \otimes \eta$, we have
$$(\xi \otimes \eta)(\xi \otimes \eta)^* = \xi \xi^* \otimes \eta \eta^*,$$
and so we have the relation

(15) $$\mathbb{V}_1 = M_m^+ \otimes M_n^+.$$

A positive semi-definite matrix in $(M_m \otimes M_n)^+$ is said to be *entangled* if it is not separable. Therefore, entanglement consists of
$$(M_m \otimes M_n)^+ \setminus M_m^+ \otimes M_n^+.$$
Recall that we have the relation
$$\mathcal{A}^+ \otimes \mathcal{B}^+ = (\mathcal{A} \otimes \mathcal{B})^+$$
if one of $C^*$-algebras $\mathcal{A}$ and $\mathcal{B}$ is commutative. This tells us that the notion of entanglement reflects non-commutative order structures in nature, and explains why there is no corresponding notion of entanglement in the classical mechanics.

The similar expression for $\mathbb{V}_s$ as (15) is also possible. It was shown in [62] that $\mathbb{V}_s = (\mathbb{P}_s)^\circ$ is the convex hull of the set
$$\left\{ \left( \sum_{i=1}^s x_i \otimes y_i \right)^* \left( \sum_{i=1}^s x_i \otimes y_i \right) \in M_m \otimes M_n : x_i \in M_m, y_i \in M_n \right\}.$$
If $s = 1$ then this says that the convex cone $\mathbb{V}_1$ is generated by $x^*x \otimes y^*y$ with $x \in M_m$ and $y \in M_n$. See also [63].

If a positive semi-definite matrix in $M_m \otimes M_n$ is of rank one in itself, then it is easy to determine if $A$ is entangled or not by definition. For example, consider the two matrices in $M_2 \otimes M_2$:

$$\begin{pmatrix} 1 & 0 & 1 & 0 \\ 0 & 0 & 0 & 0 \\ 1 & 0 & 1 & 0 \\ 0 & 0 & 0 & 0 \end{pmatrix}, \quad \begin{pmatrix} 1 & 0 & 0 & 1 \\ 0 & 0 & 0 & 0 \\ 0 & 0 & 0 & 0 \\ 1 & 0 & 0 & 1 \end{pmatrix}.$$

The first one is separable since the range vector

$$(1,0,1,0)^{\mathrm{t}} = e_1 \otimes e_1 + e_2 \otimes e_1 = (e_1 + e_2) \otimes e_1 \in \mathbb{C}^2 \otimes \mathbb{C}^2$$

is a product vector, but the second one is entangled since the range vector $(1,0,0,1)^{\mathrm{t}} \in \mathbb{C}^2 \otimes \mathbb{C}^2$ is not a product vector. If $A$ is not of rank one, it is usually very difficult to determine if $A$ is entangled or not.

It should be noted that the notion of entanglement depends on the tensor decomposition of spaces. There is an example [45] of a $6 \times 6$ matrix which is separable in $M_2 \otimes M_3$ but entangled in $M_3 \otimes M_2$.

From dual pairs $(\mathbb{V}_1, \mathbb{P}_1)$ and $(\mathbb{D}, \mathbb{T})$ together with the relation $\mathbb{D} \subset \mathbb{P}_1$, we have the following relation

(16) $$\mathbb{V}_1 \subset \mathbb{T},$$

which gives us a simple necessary condition for separability, called the PPT(positive partial transpose) criterion. The relation (16) can be seen directly, as was observed by Choi [27] and Peres [95]. Indeed, we have

(17)
$$\begin{aligned}[] [(\xi \otimes \eta)(\xi \otimes \eta)^*]^{\tau} &= [\xi\xi^* \otimes \eta\eta^*]^{\tau} \\ &= (\xi\xi^*)^{\mathrm{t}} \otimes \eta\eta^* \\ &= \bar{\xi}\bar{\xi}^* \otimes \eta\eta^* = (\bar{\xi} \otimes \eta)(\bar{\xi} \otimes \eta)^*, \end{aligned}$$

and this shows that the partial transpose of a rank one projector onto a product vector is again a rank one projector onto a product vector. The product vector $\bar{\xi} \otimes \eta$ is called the *partial conjugate* of the product vector $\xi \otimes \eta$.

By duality, it turns out that $\mathbb{P}_1[M_m, M_n] = \mathbb{D}$ if and only if $\mathbb{V}_1(M_m \otimes M_n) = \mathbb{T}$. When $m = 2$, Woronowicz [127] show that $\mathbb{V}_1 = \mathbb{T}$ if and only if $n \leq 3$, and exhibited an explicit example in $\mathbb{T} \setminus \mathbb{V}_1$ for the case of $m = 2$ and $n = 4$. This kind of example is called a *PPT entangled state* (PPTES) when it is normalized. The first example of PPTES in the case of $m = n = 3$ was given in [27]. Searching PPT entangled states is one of the main theme of this note.

The duality relation between two cones $\mathbb{V}_1$ and $\mathbb{P}_1$ gives us a characterization of separability: $A \in (M_m \otimes M_n)^+$ is separable if and only if

$$\langle A, \phi \rangle \geq 0$$

for every positive linear maps $\phi: M_m \to M_n$. Equivalently, $A \in (M_m \otimes M_n)^+$ is entangled if and only if there exists a positive linear map $\phi$ such that
$$\langle A, \phi \rangle < 0.$$
If this happens, we say that $\phi$ *detects* the entanglement $A$. A positive map which detects entanglement is said to be an *entanglement witness*, which is an another main theme of this note as well as entanglement itself. Unfortunately, the whole convex structures of the convex cone $\mathbb{P}_1$ is far from being completely understood, even in the low dimensional cases. Actually, it is now known that detecting entanglement completely is an $NP$-hard problem. See [40].

Recall that for $A \in M_m \otimes M_n$ and $\phi \in \mathcal{L}(M_m, M_n)$ the pairing $\langle A, \phi \rangle$ is nothing but
$$\langle A, \phi \rangle = \text{Tr}\,(C_\phi^{\mathrm{t}} A).$$
Therefore, we see that $A_0 \in (M_m \otimes M_n)^+$ is an entangled state if and only if there is a Hermitian matrix $W$ with the property:

(18) $\qquad\qquad \text{Tr}\,(WA_0) < 0, \qquad \text{Tr}\,(WA) \geq 0 \text{ for each } A \in \mathbb{V}_1.$

In this sense, the duality between two cones $\mathbb{V}_1$ and $\mathbb{P}_1$ is equivalent to the separability criterion given in [55] under the Jamiołkowski-Choi isomorphism. An element in the set $\mathbb{V}_s \setminus \mathbb{V}_{s-1}$ is said to have *Schmidt number s* as was introduced in [122], where the relations with $s$-positive linear maps also have been discussed.

In order to determine if a given positive semi-definite matrix in $(M_m \otimes M_n)^+$ is separable or not, it is natural to look at the range space of $A$ by the definition of separability. Assume that $A$ is separable, and write

(19) $\qquad\qquad A = z_1 z_1^* + z_2 z_2^* + \cdots + z_\iota z_\iota^*,$

with product vectors $z_i = \xi_i \otimes \eta_i \in \mathbb{C}^m \otimes \mathbb{C}^n$ for $i = 1, 2, \ldots, \iota$. If $A$ is of the form (19) with arbitrary vectors $\{z_i\}$ then it was shown [61] that the range space $\mathcal{R}A$ of $A$ coincides with the span of $\{z_i : i = 1, \ldots, \iota\}$. More generally, it is easy to see that if
$$A = P_1 + P_2 \cdots + P_\iota$$
is the sum of positive semi-definite matrices, then we have
$$\mathcal{R}A = \mathcal{R}P_1 + \mathcal{R}P_2 + \cdots + \mathcal{R}P_\iota.$$
Indeed, we have $\text{Ker}\,A \subset \text{Ker}\,P_i$ for each $i = 1, 2, \ldots \iota$, since $P_i \leq A$. Therefore, it follows that $\mathcal{R}P_i \subset \mathcal{R}A$, and so $\sum_i \mathcal{R}P_i \subset \mathcal{R}A$. The reverse inclusion is obvious. See [2]. We also have
$$A^\tau = w_1 w_1^* + w_2 w_2^* + \cdots + w_\iota w_\iota^*$$
with $w_i = \bar{\xi}_i \otimes \eta_i \in \mathbb{C}^m \otimes \mathbb{C}^n$ by (17). Therefore, it follows that

(20) $\qquad\qquad \mathcal{R}A = \text{span}\,\{\xi_i \otimes \eta_i\}, \qquad \mathcal{R}A^\tau = \text{span}\,\{\bar{\xi}_i \otimes \eta_i\}.$

Hence, we see that if $A$ is separable then there exists a family $\{\xi_i \otimes \eta_i\}$ of product vectors satisfying (20). This gives us a necessary condition for the separability, the range criterion as was shown in [57]. This is not sufficient for separability. There are examples of PPT entanglement satisfying the condition of the range criterion. See [9] for example. We will see in Section 7 how the partial converse of the range criterion works.

From now on, we identity the vector space $\mathbb{C}^m \otimes \mathbb{C}^n$ with the space $M_{m \times n}$ of all $m \times n$ matrices. Every vector $z \in \mathbb{C}^m \otimes \mathbb{C}^n$ is uniquely expressed by

$$z = \sum_{i=1}^{m} e_i \otimes z_i \in \mathbb{C}^m \otimes \mathbb{C}^n$$

with

$$z_i = \sum_{k=1}^{n} z_{ik} e_k \in \mathbb{C}^n, \qquad i = 1, 2, \ldots, m.$$

In this way, we get $z = [z_{ik}] \in M_{m \times n}$. This identification

(21) $$\sum_{i=1}^{m} e_i \otimes \left( \sum_{k=1}^{n} z_{ik} e_k \right) \longleftrightarrow [z_{ik}]$$

gives us an inner product isomorphism from $\mathbb{C}^m \otimes \mathbb{C}^n$ onto $M_{m \times n}$. Note that a product vector $\xi \otimes \bar{\eta} \in \mathbb{C}^m \otimes \mathbb{C}^n$ corresponds to the rank one matrix $\xi \eta^* \in M_{m \times n}$, and the product vector $e_i \otimes e_j \in \mathbb{C}^m \otimes \mathbb{C}^n$ corresponds to $e_{ij} \in M_{m \times n}$.

Note that the following matrix

(22) $$\begin{pmatrix} 1 & \cdot & \cdot & \cdot & 1 & \cdot & \cdot & \cdot & 1 \\ \cdot & 2 & \cdot & 1 & \cdot & \cdot & \cdot & \cdot & \cdot \\ \cdot & \cdot & \frac{1}{2} & \cdot & \cdot & \cdot & 1 & \cdot & \cdot \\ \cdot & 1 & \cdot & \frac{1}{2} & \cdot & \cdot & \cdot & \cdot & \cdot \\ 1 & \cdot & \cdot & \cdot & 1 & \cdot & \cdot & \cdot & 1 \\ \cdot & \cdot & \cdot & \cdot & \cdot & 2 & \cdot & 1 & \cdot \\ \cdot & \cdot & 1 & \cdot & \cdot & \cdot & 2 & \cdot & \cdot \\ \cdot & \cdot & \cdot & \cdot & \cdot & 1 & \cdot & \frac{1}{2} & \cdot \\ 1 & \cdot & \cdot & \cdot & 1 & \cdot & \cdot & \cdot & 1 \end{pmatrix}$$

belongs to the cone $\mathbb{T}$. Note also that the range is the 4-dimensional space spanned by

$$e_1 \otimes e_1 + e_2 \otimes e_2 + e_3 \otimes e_3$$

and

$$\sqrt{2} e_1 \otimes e_2 + \frac{1}{\sqrt{2}} e_2 \otimes e_1, \qquad \sqrt{2} e_2 \otimes e_3 + \frac{1}{\sqrt{2}} e_3 \otimes e_2, \qquad \sqrt{2} e_3 \otimes e_1 + \frac{1}{\sqrt{2}} e_1 \otimes e_3.$$

It is easy to see that the corresponding 4-dimensional subspace of $M_{3 \times 3}$ spanned by

$$\begin{pmatrix} 1 & 0 & 0 \\ 0 & 1 & 0 \\ 0 & 0 & 1 \end{pmatrix}, \quad \begin{pmatrix} 0 & \sqrt{2} & 0 \\ \frac{1}{\sqrt{2}} & 0 & 0 \\ 0 & 0 & 0 \end{pmatrix}, \quad \begin{pmatrix} 0 & 0 & 0 \\ 0 & 0 & \sqrt{2} \\ 0 & \frac{1}{\sqrt{2}} & 0 \end{pmatrix}, \quad \begin{pmatrix} 0 & 0 & \frac{1}{\sqrt{2}} \\ 0 & 0 & 0 \\ \sqrt{2} & 0 & 0 \end{pmatrix}$$

has no rank one matrices, which implies that the matrix in (22) is entangled. This is the first example of $3 \otimes 3$ PPTES given by Choi [27].

We say that a subspace of $\mathbb{C}^m \otimes \mathbb{C}^n$ is *completely entangled* if it has no nonzero product vector. Note that a positive semi-definite matrix with the completely entangled range space is never separable. It is known [68] that the maximal dimension of completely entangled subspaces in $\mathbb{C}^m \otimes \mathbb{C}^n$ is given by

$$p = (m-1) \times (n-1),$$

and the set of $p$-dimensional subspaces that contain product vectors is of codimension one in the set of all $p$-dimensional subspaces in $\mathbb{C}^m \otimes \mathbb{C}^n$. Furthermore, generic $(p+1)$-dimensional subspaces contain exactly $\binom{m+n-2}{n-1}$ lines induced by product vectors. See also [15], [93], [125] and [126].

We refer to the book [12] for another criteria for separability as well as more general aspects of the theory of entanglement. See also [39] and [60].

## 4. Faces for completely positive maps

A convex subset $F$ of a convex set $C$ is said to be a face of $C$ if the following condition

$$x, y \in C, \ (1-t)x + ty \in F \text{ for some } t \in (0,1) \implies x, y \in F$$

holds. An extreme point is a face consisting of a single point. If a ray $\{\lambda x : \lambda \geq 0\}$ is a face of a convex cone $C$ then it is called an *extreme ray*, and we say that $x$ generates an extreme ray.

A point $x_0$ of a convex set $C$ is said to be an *interior point* of $C$ if for any $x \in C$ there is $t > 1$ such that $(1-t)x + tx_0 \in C$. Geometrically, a point $x_0$ is an interior point of $C$ if and only if the line segment from any point of $C$ to $x_0$ may be extended inside of $C$. If $C$ is a convex subset of a finite dimensional space then the set $\text{int} \, C$ of all interior points of $C$ is nothing but the relative topological interior of $C$ with respect to the affine manifold generated by $C$. Note that $\text{int} \, C$ is never empty for any nonempty convex set $C$. If one interior point $y$ of $C$ is known, then it is easy to see that $x_0 \in C$ is an interior point of $C$ if and only if there is $t > 1$ such that $(1-t)y + tx_0 \in C$. See [77]. It is known that a convex set is partitioned into the interiors of faces. See [99], Theorem 18.2. Therefore, we see that a point $x$ of a convex set gives rise to a unique face in which $x$ is an interior point. This is the smallest face containing $x$. A point of $C$ is said to be a *boundary point* if it is not an interior point, and we denote by $\partial C$ the set of all boundary points of $C$.

For a subset $F$ of a closed convex cone $C$ of $X$, we define the subset $F'$ of $C^\circ$ by

$$F' = \{y \in C^\circ : \langle x, y \rangle = 0 \text{ for each } x \in F\} \subset C^\circ \subset Y.$$

It is then clear that $F'$ is a face of $C^\circ$, which is said to be the *dual face* of $F$. If $F$ is a face with an interior point $x_0$ then we see that

$$F' = \{y \in C^\circ : \langle x_0, y \rangle = 0\}.$$

Similarly, we also define the dual face $G'$ of $C$ for a face $G$ of $C^\circ$. We say that $F \subset C$ is an *exposed face* if it is a dual face. It is easy to see that a face $F$ is exposed if and only if $F = F''$. If $\{x\}$ is a singleton then $\{x\}'$ will be denoted just by $x'$.

Now, we pay attention to the dual pair $(\mathbb{P}_{m \wedge n}, \mathbb{V}_{m \wedge n})$, and proceed to determine the dual faces. For a vector $z = \sum_{i=1}^{m} e_i \otimes z_i \in \mathbb{C}^m \otimes \mathbb{C}^n$ and a completely positive map $\phi_V \in \mathbb{P}_{m \wedge n}$ with $V = [v_{ik}] \in M_{m \times n}$, we have

$$zz^* = \sum_{i,j=1}^{m} e_{ij} \otimes z_i z_j^* \in M_m \otimes M_n.$$

Therefore, it follows that

$$\langle zz^*, \phi_V \rangle = \sum_{i,j=1}^{m} \langle z_i z_j^*, V_i^* V_j \rangle$$

by the relation (3), where $V_i$ is a row vector which is the $i$th row of the matrix $V \in M_{m \times n}$, and $z_i$ is the the column vector which is the $i$th block of $z \in \mathbb{C}^m \otimes \mathbb{C}^n$. We see that

$$\langle z_i z_j^*, V_i^* V_j \rangle = \mathrm{Tr}\,(V_i^* V_j (z_i z_j^*)^{\mathrm{t}}) = \mathrm{Tr}\,(V_i^* V_j \bar{z}_j z_i^{\mathrm{t}}) = V_j \bar{z}_j \mathrm{Tr}\,(V_i^* z_i^{\mathrm{t}}) = (V_j | z_j)(z_i | V_i),$$

where $(\ |\ )$ denotes the inner product of the space $\mathbb{C}^n$. Therefore, it follows that

(23) $$\langle zz^*, \phi_V \rangle = \left|\sum_{i=1}^{m}(z_i\,|\,V_i)\right|^2 = |(z\,|\,V)|^2$$

if we identity $z$ as an $m \times n$ matrix by (21), where $(\ |\ )$ in the right-hand side denotes the inner product of the space $M_{m \times n}$.

For a given completely positive map $\phi_\mathcal{V}$ with a subset $\mathcal{V}$ of $M_{m \times n}$, we see that $A \in \mathbb{V}_{m \wedge n}$ belongs to the dual face of $\phi_\mathcal{V}$ if and only if the range space of $A$ is orthogonal to the span $D$ of $\mathcal{V}$. Therefore, every exposed face of the cone $\mathbb{V}_{m \wedge n}$ is of the form

$$\tau_{D^\perp} := \{A \in \mathbb{V}_{m \wedge n} : \mathcal{R}A \subset D^\perp\},$$

for a subspace $D$ of $M_{m \times n} = \mathbb{C}^m \otimes \mathbb{C}^n$ by the identification (21). Note that every face of the convex cone of all positive semi-definite matrices is of this form. See [10]. It is clear that the following relation

(24) $$\mathrm{int}\,\tau_{D^\perp} = \{A \in \mathbb{V}_{m \wedge n} : \mathcal{R}A = D^\perp\}$$

holds.

It is also apparent that the dual face of $\tau_{D^\perp}$ is given by

$$\sigma_D := \{\phi_\mathcal{V} : \mathcal{V} \subset D\}.$$

We show that every face of the cone $\mathbb{P}_{m \wedge n}$ is of this form for a subspace $D$ of $M_{m \times n} = \mathbb{C}^m \otimes \mathbb{C}^n$, and so it is exposed. To do this, let $F$ be the smallest face of $\mathbb{P}_{m \wedge n}$ containing the map $\phi_\mathcal{V}$. It suffices to show the following:

$$\mathrm{span}\,\mathcal{W} \subset \mathrm{span}\,\mathcal{V} \implies \phi_\mathcal{W} \in F.$$

We may assume that $\mathcal{V} = \{V_k : k = 1, 2, \ldots, s\}$ and $\mathcal{W} = \{W_\ell : \ell = 1, 2, \ldots, r\}$ are linearly independent. Write

$$W_\ell = \sum_{k=1}^{s} a_{\ell k} V_k, \qquad \ell = 1, 2, \ldots, r.$$

Then we have

$$\phi_{\mathcal{W}}(X) = \sum_{\ell=1}^{r} \left(\sum_{k=1}^{s} a_{\ell k} V_k\right)^* X \left(\sum_{j=1}^{s} a_{\ell j} V_j\right) = \sum_{k,j=1}^{s} \left(\sum_{\ell=1}^{r} \overline{a_{\ell k}} a_{\ell j}\right) V_k^* X V_j.$$

We write $A$ for the $r \times s$ matrix whose $(\ell, k)$-entry is $a_{\ell k}$. Then there is $t > 1$ such that $(1-t)A^*A + t(I_m \otimes I_n)$ is positive semi-definite, which will be denoted by $B^*B$, with an $s \times s$ matrix $B = [b_{\ell k}]$. Then we have

$$[(1-t)\phi_{\mathcal{W}} + t\phi_{\mathcal{V}}](X) = \sum_{k,j=1}^{s} \left(\sum_{\ell=1}^{s} \overline{b_{\ell k}} b_{\ell j}\right) V_k^* X V_j = \sum_{\ell=1}^{s} \left(\sum_{k=1}^{s} b_{\ell k} V_k\right)^* X \left(\sum_{j=1}^{s} b_{\ell j} V_j\right),$$

and so it follows that $\phi := (1-t)\phi_{\mathcal{W}} + t\phi_{\mathcal{V}}$ is completely positive. This shows that $\phi_{\mathcal{V}}$ is a nontrivial convex combination of $\phi_{\mathcal{W}}$ and $\phi \in \mathbb{P}_{m \wedge n}$. Since $\phi_{\mathcal{V}} \in F$, we conclude that $\phi_{\mathcal{W}} \in F$. It is apparent that

(25) $$\operatorname{int} \sigma_D = \{\phi_{\mathcal{V}} : \operatorname{span} \mathcal{V} = D\}.$$

We can summarize our discussion as in [79], where the convex set of all unital completely positive maps has been considered. See also [11].

**Theorem 4.1.** *Every face of the cone $\mathbb{P}_{m \wedge n}$ is exposed, and the correspondence*

$$D \mapsto \sigma_D$$

*defines a lattice isomorphism from the complete lattice of all subspaces of $M_{m \times n}$ onto the complete lattice of all faces of the cone $\mathbb{P}_{m \wedge n}$. We also have*

$$(\sigma_D)' = \tau_{D^\perp},$$

*with respect to the duality between $\mathbb{P}_{m \wedge n}$ and $\mathbb{V}_{m \wedge n}$.*

Especially, we see that the ray generated by $\phi_V$ is an exposed face of $\mathbb{P}_{m \wedge n}$, which is automatically generates an extremal ray. It is known that $\phi_V$ also generates an exposed ray of the much bigger cone $\mathbb{P}_1$. See [131] and [90].

As for the dual pair $(\mathbb{P}^{m \wedge n}, \mathbb{V}^{m \wedge n})$, we also have

$$\langle (zz^*)^\tau, \phi^V \rangle = \langle zz^*, \phi^V \circ \operatorname{tp} \rangle = \langle zz^*, \phi_V \rangle = |(z \,|\, V)|^2,$$

and the same argument holds.

**Theorem 4.2.** *Every face of the cone $\mathbb{P}^{m \wedge n}$ is exposed, and the correspondence*

$$E \mapsto \sigma^E := \{\phi^\mathcal{V} : \mathcal{V} \subset D\}$$

defines a lattice isomorphism from the complete lattice of all subspaces of $M_{m\times n}$ onto the complete lattice of all faces of the cone $\mathbb{P}^{m\wedge n}$. We have

$$(\sigma^E)' = \tau^{E^\perp} := \{A^\tau \in \mathbb{V}^{m\wedge n} : \mathcal{R}A \subset E^\perp\},$$

with respect to the duality between $\mathbb{P}^{m\wedge n}$ and $\mathbb{V}^{m\wedge n}$. We also have

(26) $$\operatorname{int} \sigma^E = \{\phi^\mathcal{V} : \operatorname{span} \mathcal{V} = D\}.$$

## 5. Boundary structures for positive maps

Although the whole facial structures of the cone $\mathbb{P}_s$ is still mysterious for $s < m \wedge n$ as well as for $s = 1$, it is possible to characterize the boundaries of these cones using the duality between $\mathbb{P}_s$ and $\mathbb{V}_s$, since we know all extreme rays of the cone $\mathbb{V}_s$ by definition. Note that the boundary of a convex set consists of maximal faces.

For a product vector $z = \xi \otimes \eta \in \mathbb{C}^m \otimes \mathbb{C}^n$, we have

(27) $$\langle zz^*, \phi \rangle = \langle \xi\xi^* \otimes \eta\eta^*, \phi \rangle = \operatorname{Tr}(\phi(\xi\xi^*)\bar\eta\bar\eta^*) = (\phi(\xi\xi^*)\bar\eta|\bar\eta).$$

This relation shows the following:

$$\phi \in \mathbb{P}_1, \ \langle A, \phi \rangle = 0 \text{ for each } A \in \mathbb{V}_1 \implies \phi = 0.$$

Let $X$ and $Y$ be finite-dimensional normed spaces, which are dual each other with respect to a bilinear pairing $\langle\,,\,\rangle$, as before. We also assume that $C$ is a closed convex cone of $X$ on which the pairing is *non-degenerate*, that is,

(28) $$x \in C, \ \langle x, y \rangle = 0 \text{ for each } y \in C^\circ \implies x = 0.$$

By the compactness argument, we see that this assumption guarantees the existence of a point $\eta \in C^\circ$ with the property:

(29) $$x \in C, \ x \neq 0 \implies \langle x, \eta \rangle > 0,$$

which is seemingly stronger than (28). As an another immediate consequence of (28), we also have

(30) $$F \text{ is a face of } C, \ F' = C^\circ \implies F = \{0\}.$$

**Proposition 5.1.** *Let $X$ and $Y$ be finite-dimensional normed spaces with a non-degenerate bilinear pairing $\langle\,,\,\rangle$ on a closed convex cone $C$ in $X$. For a given point $y \in C^\circ$, the following are equivalent:*

  (i) *$y$ is an interior point of $C^\circ$.*
  (ii) *$\langle x, y \rangle > 0$ for each nonzero $x \in C$.*

*Proof.* If $y$ is an interior point of $C^\circ$ then we may take $t \in [0, 1)$ and $z \in C^\circ$ such that $y = (1-t)\eta + tz$, where $\eta \in C^\circ$ is a point with the property (29). Then we see that

$$\langle x, y \rangle = (1-t)\langle x, \eta \rangle + t\langle x, z \rangle > 0$$

for each nonzero $x \in C$. Now, we assume (ii), and take an arbitrary point $z \in C^\circ$. Put $C_\epsilon = \{x \in C : \|x\| = \epsilon\}$. Then since $C_1$ is compact, $\alpha = \sup\{\langle x, z \rangle : x \in C_1\}$ is finite, and we see that $\langle x, z \rangle \leq 1$ for each $x \in C_{1/\alpha}$. We also take $\delta$ with $0 < \delta < 1$ such that $\langle x, y \rangle \geq \delta$ for each $x \in C_{1/\alpha}$. Put

$$w = \left(1 - \frac{1}{1-\delta}\right) z + \frac{1}{1-\delta} y.$$

Then we see that $\langle x, w \rangle \geq 0$ for each $x \in C_{1/\alpha}$, and so $w \in C^\circ$. Since $z$ was an arbitrary point of $C^\circ$ and $\frac{1}{1-\delta} > 1$, we see that $y$ is an interior point of $C^\circ$. □

A typical interior point of the cone $\mathbb{P}_s[M_m, M_n]$ is the trace map

$$\mathrm{Tr} : X \mapsto \mathrm{Tr}(X) I_n, \qquad X \in M_n,$$

whose Choi matrix $C_{\mathrm{Tr}}$ is nothing but the identity matrix of $M_m \otimes M_n$. Since every nontrivial face lies on the boundary, Proposition 5.1 tells us the following:

(31) $\qquad\qquad F$ is a face of $C^\circ$, $F' = \{0\} \implies F = C^\circ$.

Indeed, if $F$ is a nontrivial face of $C^\circ$ then we can take a nonzero $y \in \mathrm{int}\, F \subset \partial C^\circ$. Then we have $\langle x, y \rangle = 0$ for a nonzero $x \in C$, and so $x \in F'$. This shows that $F$ is nonzero.

We say that a point of a closed convex cone is *extreme* (respectively *exposed*) if it generates an extreme ray (respectively an exposed ray). An exposed point is automatically extreme. We note that every element of the cone $C$ is the convex sum of extreme points of $C$, and every extreme point is the limit of exposed points by Straszewicz's Theorem (see [99], Theorem 18.6). Therefore, we have the following:

**Proposition 5.2.** *Let $X$ and $Y$ be finite-dimensional normed spaces with a bilinear pairing. For a convex cone $C$ in $X$ and $y \in Y$, the following are equivalent:*

(i) $y \in C^\circ$.
(ii) $\langle x, y \rangle \geq 0$ for every extreme point of $x$ of $C$.
(iii) $\langle x, y \rangle \geq 0$ for every exposed point of $x$ of $C$.

We say that $L$ is a *minimal exposed face* if it is an exposed face which is minimal among all exposed faces. If $L$ is a minimal exposed face of the cone $C$ then $L'$ is a maximal face of $C^\circ$. To see this, let $F$ be a face of $C^\circ$ such that $F \supset L'$. Then we have

$$L = L'' \supset F'.$$

Since $F'$ is an exposed face, we have $F' = \{0\}$ or $F' = L$. If $F' = \{0\}$ then $F = C^\circ$ by (31). If $F' = L$ then $F \subset F'' = L'$, which implies $F = L'$. This shows that $L'$ is a maximal face. We proceed to show that every maximal face of $C^\circ$ is of the form $L'$ for a minimal exposed face $L$ of $C$. If $F$ is a maximal face of $C^\circ$ then $F$ lies on the boundary of $C^\circ$. If we take an interior point $y_0$ of $F$ then there is $x_0 \in C$ such that $\langle x_0, y_0 \rangle = 0$ by Proposition 5.1. Take the face $L$ in which $x_0$ is an interior point. Then we see that $y_0 \in L' \cap \mathrm{int}\, F$, from which we infer that $F \subset L'$. Because $L' \subsetneq C^\circ$ by (30), we have $F = L' = (L'')'$. Especially, $F$ is exposed by the exposed face $L''$, which is the smallest exposed face containing $x_0$. From the maximality of $F$, it is apparent that $L''$ is minimal among all exposed faces. If $L'_1 = L'_2 = F$ for exposed faces $L_1$ and $L_2$, then we have

$$L_1 = L''_1 = F' = L''_2 = L_2,$$

and so, we see that every maximal face $F$ is the dual face of a unique minimal exposed face $L$.

**Proposition 5.3.** *Let $X$ and $Y$ be finite-dimensional normed spaces with a non-degenerate bilinear pairing $\langle\,,\,\rangle$ on a closed convex cone $C$ in $X$. If $L$ is a minimal exposed face of $C$ then $L'$ is a maximal face of $C^\circ$. Conversely, every maximal face of $C^\circ$ is the dual face of a unique minimal exposed face of $C$.*

Note that an exposed ray is automatically a minimal exposed face. The converse is not true in general. Since every convex cone has an exposed ray, every minimal exposed face has an exposed ray in itself, but this ray need not to be exposed in the whole convex cone.

If $y$ is a boundary point of $C^\circ$ then it is an element of a maximal face $F$, which is the dual face of an interior point $x$ of a minimal exposed face $L$ of $C$. This means $\langle x, y \rangle = 0$. Therefore, we have the following extension of Proposition 5.1. It is clear that the statement (ii) of the following is equivalent to (ii) of Proposition 5.1, since every point of $C$ is the convex sum of extreme points of $C$.

**Proposition 5.4.** *Let $X$ and $Y$ be finite-dimensional normed spaces with a non-degenerate bilinear pairing $\langle\,,\,\rangle$ on a closed convex cone $C$ in $X$. For a given point $y \in C^\circ$, the following are equivalent:*

(i) *$y$ is an interior point of $C^\circ$.*
(ii) *$\langle x, y \rangle > 0$ for every extreme point $x$ of $C$.*
(iii) *$\langle x, y \rangle > 0$ for an interior point $x$ of $L$, for every minimal exposed face $L$ of $C$.*

Now, we apply the above discussion to the dual pair $(\mathbb{V}_s, \mathbb{P}_s)$. Note that every extreme ray of the cone $\mathbb{V}_s$ is generated by $zz^*$ for an $s$-simple vector $z \in \mathbb{C}^m \otimes \mathbb{C}^n$ by the definition of the cone $\mathbb{V}_s$. Since this ray is already an exposed face of the bigger cone $\mathbb{V}_{m \wedge n}$, it is apparent that every extremal ray of the cone $\mathbb{V}_s$ is exposed. This means that a face of the cone $\mathbb{V}_s$ is an exposed ray if and only if it is a minimal exposed face. Therefore, we may apply Proposition 5.3 to see the following:

**Theorem 5.5.** *For each $s$-simple vector $z \in \mathbb{C}^m \otimes \mathbb{C}^n$, the set*

$$\{\phi \in \mathbb{P}_s : \langle zz^*, \phi \rangle = 0\} \qquad (\text{respectively } \{\phi \in \mathbb{P}^s : \langle (zz^*)^\tau, \phi \rangle = 0\})$$

*is a maximal face of $\mathbb{P}_s$ (respectively $\mathbb{P}^s$). Conversely, every maximal face of $\mathbb{P}_s$ (respectively $\mathbb{P}^s$) arises in this form for a unique $s$-simple vector $z \in \mathbb{C}^m \otimes \mathbb{C}^n$ up to scalar multiples.*

**Corollary 5.6.** *A map $\phi \in \mathbb{P}_s$ is on the boundary of the cone $\mathbb{P}_s$ if and only if there exists an $s$-simple vector $z \in \mathbb{C}^m \otimes \mathbb{C}^n$ such that $\langle zz^*, \phi \rangle = 0$.*

See [78] for an another description for maximal faces of the cone $\mathbb{P}_s$ which is equivalent to Theorem 5.5. The most interesting case is when $s = 1$. In this case, we see by (27) that every maximal face is of the form

$$\{\phi \in \mathbb{P}_1 : (\phi(\xi\xi^*)\eta \,|\, \eta) = 0\},$$

for a product vector $\xi \otimes \eta \in \mathbb{C}^m \otimes \mathbb{C}^n$. Therefore, we see that $\phi \in \mathbb{P}_1$ is on the boundary of the cone $\mathbb{P}_1$ if and only if there is nonzero $\xi \in \mathbb{C}^m$ such that $\phi(\xi\xi^*)$ is singular in $M_n$. Using this, it is possible to construct a join homomorphism from the lattice of all faces of the cone $\mathbb{P}_1$ into the lattice of all join homomorphisms between the lattices of all subspaces of $\mathbb{C}^m$ and $\mathbb{C}^n$, respectively. See [77]. We also see that maximal faces of the cone $\mathbb{P}_1$ are parameterized by the product of two complex projective spaces. It is known [77] that any two maximal faces of the cone $\mathbb{P}_1$ are affine isomorphic.

Note that an extreme point of the cone $\mathbb{D}$ is either $\phi_V$ or $\phi^V$ for a matrix $V$. So, these are only candidates of exposed decomposable maps in the cone $\mathbb{P}_1$. It is known [131] that they are always extreme in $\mathbb{P}_1$, and exposed in the cone $\mathbb{P}_1$ if the rank of $V$ is one or full. More recently, it was shown in [90] that $\phi_V$ is always exposed.

Among positive maps in Theorem 1.2, consider the maps with the following condition

(32) $$0 < a < 1, \qquad a+b+c = 2, \qquad bc = (1-a)^2.$$

Motivated by a parametrization [36] for those cases, it was shown in [49] that $\Phi[a,b,c]$ is an exposed positive linear map whenever the conditions (32) holds. See also [35] and [48].

Even though every maximal face is exposed in general, it should be noted that there is a face of $\mathbb{P}_1$ which is not exposed. Indeed, if we slice the convex body for $\mathbb{P}_1$ in Theorem 1.2 with the hyperplane $a+b+c = 2$, then it is clear by the two-dimensional picture that the Choi map $\phi = \Phi[1, 0, 1]$ is not exposed. See Figure 2. It is worthwhile to calculate the dual face of the Choi map. To do this, we first find all vectors $\xi \in \mathbb{C}^3$ such that $\phi(\xi\xi^*)$ is singular, and find null vectors $\eta \in \mathbb{C}^3$ of $\phi(\xi\xi^*)$. Then the dual face will be generated by those rank one projectors in $M_3 \otimes M_3$ onto product vector $\xi \otimes \bar{\eta}$ by (27).

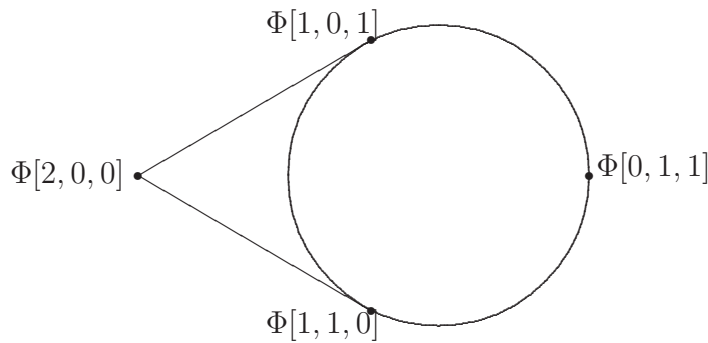

FIGURE 2. The region representing positive linear maps in (4) on the plane $a+b+c = 2$. Points on the parts of the circle are the intersection points of the hyperbola and the straight line in Figure 1: $a = \frac{1}{2}$. Points on the line segments are intersection points of the line and axes in Figure 1: $a = \frac{5}{4}$, $a = \frac{3}{2}$.

By a direct calculation, we see that $\phi(\xi\xi^*)$ is singular if and only if $\xi$ is one of the following vectors

$$\xi_1 = (1, 0, 0), \qquad \xi_2 = (0, 1, 0), \qquad \xi_3 = (0, 0, 1), \qquad \xi_4 = (e^{ia}, e^{ib}, e^{ic}),$$

and the corresponding null spaces are generated by

$$\eta_1 = (0, 0, 1), \qquad \eta_2 = (1, 0, 0), \qquad \eta_3 = (0, 1, 0), \qquad \eta_4 = (e^{ia}, e^{ib}, e^{ic}),$$

respectively. If we identify $\xi_i \otimes \bar{\eta}_i$ with the rank one matrix $\xi_i \eta_i^*$ by (21), then we see that the projector onto $\xi \otimes \bar{\eta}$ belongs to the dual face of the Choi map if and only $\xi\eta^*$ is one of the following matrices:

$$\begin{pmatrix} 0 & 0 & 1 \\ 0 & 0 & 0 \\ 0 & 0 & 0 \end{pmatrix}, \quad \begin{pmatrix} 0 & 0 & 0 \\ 1 & 0 & 0 \\ 0 & 0 & 0 \end{pmatrix}, \quad \begin{pmatrix} 0 & 0 & 0 \\ 0 & 0 & 0 \\ 0 & 1 & 0 \end{pmatrix}, \quad \begin{pmatrix} 1 & \alpha & \bar{\gamma} \\ \bar{\alpha} & 1 & \beta \\ \gamma & \bar{\beta} & 1 \end{pmatrix},$$

where $\alpha\beta\gamma = 1$ with $|\alpha| = |\beta| = |\gamma| = 1$. We show that those matrices span the following 7-dimensional space:

$$(33) \qquad D = \{[a_{ij}] \in M_3 : a_{11} = a_{22} = a_{33}\}.$$

It is clear that every matrix $[a_{ij}]$ in $D$ has the relation $a_{11} = a_{22} = a_{33}$, and so the dimension of $D$ is at most 7. We see [22] that the following four matrices

$$\begin{pmatrix} 1 & 1 & 1 \\ 1 & 1 & 1 \\ 1 & 1 & 1 \end{pmatrix}, \quad \begin{pmatrix} 1 & -1 & 1 \\ -1 & 1 & -1 \\ 1 & -1 & 1 \end{pmatrix}, \quad \begin{pmatrix} 1 & 1 & -1 \\ 1 & 1 & -1 \\ -1 & -1 & 1 \end{pmatrix}, \quad \begin{pmatrix} 1 & -1 & -1 \\ -1 & 1 & 1 \\ -1 & 1 & 1 \end{pmatrix}$$

together with $e_{21}, e_{32}, e_{13}$ are linearly independent rank one matrices belonging to $D$.

By Figure 2 for the convex body sliced by the plane $a + b + c = 2$, it is also clear that $\Phi[2, 0, 0]$ belongs to the bidual face of the Choi map. Recall the relation

$$\Phi[2, 0, 0] = \phi_{V_1} + \phi_{V_2} + \phi_{V_3}$$

with

$$V_1 = e_{11} - e_{22}, \qquad V_2 = e_{22} - e_{33}, \qquad V_3 = e_{33} - e_{11},$$

and note that $V_i$ is orthogonal to the space $D$ for each $i = 1, 2, 3$. It is easy to see that that a completely positive map $\phi_V$ lies in the bidual cone of the Choi map if and only if $V$ is orthogonal to the space $D$. This will be clarified in general situations later and exploited to study the notion of entanglement witnesses. See the relation (52) and Proposition 8.3.

Woronowicz [129] kindly showed the author that if a positive map $\phi \in \mathbb{P}_1(M_m, M_n)$ satisfies the following two conditions

- $\phi$ is irreducible; $\{x \in M_n : \phi(a)x = x\phi(a) \text{ for each } a \in M_m\} = \mathbb{C}I_n$,
- $\dim \mathrm{span}\, \{a \otimes h : M_m^+ \otimes \mathbb{C}^n : \phi(a)h = 0\} = n \times (m^2 - 1)$,

then it is exposed. Note that the second condition appears in Theorem 3.3 of [128] in the context of the notion of non-extendability. Very recently, examples satisfying these conditions have been found in [104]. Another examples of indecomposable exposed maps can be found in [30].

It is clear that the discussions in this section might be applied to describe the maximal faces of the cone $\mathbb{V}_1$, which would give us the boundary structures between entanglement and separable ones. Nevertheless, there is no known criterion to determine if a separable state is on the boundary of the cone $\mathbb{V}_1$ or not. We refer to [1], [2] and [22] for facial structures of the cone $\mathbb{V}_1$. See also [52] for a recent progresses.

## 6. Faces for decomposable maps and partial transposes

In this section, we pay attention to the duality between the cone $\mathbb{D}$ and the cone $\mathbb{T}$, and describe their facial structures. Recall that the cone $\mathbb{D}$ is the convex hull generated by the cones $\mathbb{P}_{m \wedge n}$ and $\mathbb{P}^{m \wedge n}$, and the cone $\mathbb{T}$ is the intersection of the cones $\mathbb{V}_{m \wedge n}$ and $\mathbb{V}^{m \wedge n}$. We begin with the general situation.

Let $C_1$ and $C_2$ be closed convex cones of a normed vector space $X$. If $F$ is a face of the cone $C_1 + C_2$ generated by $C_1$ and $C_2$ then it is easy to see that $F_i = F \cap C_i$ is a face of $C_i$ for $i = 1, 2$ and the identity

$$F = F_1 + F_2$$

holds. Therefore, every face of the cone $C_1 + C_2$ is determined by a pair of faces. It should be noted that different pairs may give rise to the same face. But, it is clear that if we assume the condition

$$(34) \qquad (F_1 + F_2) \cap C_i = F_i, \qquad i = 1, 2,$$

then the pair $(F_1, F_2)$ generating $F$ is uniquely determined.

On the other hand, if $F_i$ is a face of the cone $C_i$ for $i = 1, 2$ then $F_1 \cap F_2$ is a face of $C_1 \cap C_2$. Conversely, every face $F$ of the cone $C = C_1 \cap C_2$ associates with a unique pair $(F_1, F_2)$ of faces of $C_1$ and $C_2$, respectively, with the properties

$$(35) \qquad F = F_1 \cap F_2, \qquad \text{int } F \subset \text{int } F_1, \qquad \text{int } F \subset \text{int } F_2.$$

To see this, take an interior point $x$ of $F$ in $C_1 \cap C_2$. If we take the face $F_i$ of $C_i$ with $x \in \text{int } F_i$ for $i = 1, 2$ then we have

$$x \in \text{int } F_1 \cap \text{int } F_2 \subset \text{int } (F_1 \cap F_2).$$

Since $F_1 \cap F_2$ is a face of $C$, we conclude that $F = F_1 \cap F_2$. The uniqueness is clear, because every convex set is decomposed into the interiors of faces.

Now, we proceed to consider the duality. Let $F_i$ be a face of the convex cone $C_i$, for $i = 1, 2$, satisfying the conditions in (34) such that $F_1 + F_2$ is a face of the cone $C_1 + C_2$. It is easy to see that

$$(36) \qquad (F_1 + F_2)' = F_1' \cap F_2',$$

where it should be noted that the dual faces should be taken in the corresponding duality. For example, $(F_1 + F_2)'$ is the set of all $y \in C^\circ = C_1^\circ \cap C_2^\circ$ such that $\langle x, y \rangle = 0$ for each $x \in F_1 + F_2$. On the other hand, $F_i'$ is the set of all $y \in C_i^\circ$ such that $\langle x, y \rangle = 0$ for each $x \in F_i$ for $i = 1, 2$. Analogously, if $F_i$ is a face of $C_i$ satisfying (35) then we have

$$(37) \qquad (F_1 \cap F_2)' = F_1' + F_2'.$$

From the easy inclusion $F_i' \subset (F_1 \cap F_2)'$, one direction comes out. For the reverse inclusion, let $y \in (F_1 \cap F_2)'$. Since $y \in (C_1 \cap C_2)^\circ = C_1^\circ + C_2^\circ$, we may write $y = y_1 + y_2$ with $y_i \in C_i^\circ$ for $i = 1, 2$. We also take an interior point $x$ of $F_1 \cap F_2$. Then we have $x \in \operatorname{int} F_i \subset C_i$ by (35), and so $\langle x, y_i \rangle \geq 0$ for $i = 1, 2$. From the relation

$$0 = \langle x, y \rangle = \langle x, y_1 \rangle + \langle x, y_2 \rangle,$$

we conclude that $\langle x, y_i \rangle = 0$. Since $x$ is an interior point of $F_i$, we see that $y_i \in F_i'$ for $i = 1, 2$, and $y \in F_1' + F_2'$.

Now, we apply the above results to the following two cones

$$\mathbb{D} = \mathbb{P}_{m \wedge n} + \mathbb{P}^{m \wedge n}, \qquad \mathbb{T} = \mathbb{V}_{m \wedge n} \cap \mathbb{V}^{m \wedge n}.$$

We say that a pair $(D, E)$ of subspaces of $M_{m \times n}$ is a *decomposition pair* if $\sigma_D + \sigma^E$ is a face of $\mathbb{D}$ and the condition

$$(\sigma_D + \sigma^E) \cap \mathbb{P}_{m \wedge n} = \sigma_D, \qquad (\sigma_D + \sigma^E) \cap \mathbb{P}^{m \wedge n} = \sigma^E$$

holds. This is an another expression of (34). Then every face of the cone $\mathbb{D}$ is of the form

$$\sigma(D, E) := \sigma_D + \sigma^E$$

for a unique decomposition pair $(D, E)$ of subspaces, as was seen in [81]. We use the notation $\sigma(D, E)$ only when $(D, E)$ is a decomposition pair. On the other hand, we say that a pair $(D, E)$ is an *intersection pair* if the condition

$$\operatorname{int}(\tau_D \cap \tau^E) \subset \operatorname{int} \tau_D \cap \operatorname{int} \tau^E$$

holds, as is in (35). Note that the reverse inclusion holds always. Then every face of the cone $\mathbb{T}$ is of the form

$$\tau(D, E) := \tau_D \cap \tau^E$$

for a unique intersection pair. The notation $\tau(D, E)$ will be also used only when $(D, E)$ is an intersection pair. The relations (36) and (37) may be translated into the following:

$$(38) \qquad \sigma(D, E)' = \tau_{D^\perp} \cap \tau^{E^\perp}, \qquad \tau(D, E)' = \sigma_{D^\perp} + \sigma^{E^\perp}.$$

We will see that if $(D, E)$ is an intersection pair then $(D^\perp, E^\perp)$ is a decomposition pair. It should be noted that $(D^\perp, E^\perp)$ is not necessarily an intersection pair, even though $(D, E)$ is a decomposition pair.

Now, we determine exposed faces among all faces $\sigma(D, E)$ of the cone $\mathbb{D}$, and use this to show that every face of the cone $\mathbb{T}$ is exposed, as in [46]. Note that subspaces $D$ and $E$ of $M_{m \times n}$ may be considered as subspaces of $\mathbb{C}^m \otimes \mathbb{C}^n$ by the correspondence (21).

**Lemma 6.1.** *Suppose that $\sigma(D, E)$ is an exposed face of $\mathbb{D}$ and $\sigma(D, E) = A'$ for $A \in \mathbb{T}$ then we have $\mathcal{R}A = D^\perp$ and $\mathcal{R}A^\tau = E^\perp$.*

*Proof.* First of all, the relation
$$A \in A'' = \sigma(D, E)' = \tau_{D^\perp} \cap \tau^{E^\perp}$$
implies that $\mathcal{R}A \subset D^\perp$ and $\mathcal{R}A^\tau \subset E^\perp$. For the reverse inclusion, let $V \in M_{m \times n}$ with $V \in (\mathcal{R}A)^\perp$, and write $A = \sum_i z_i z_i^*$ with $z_i \in M_{m \times n} = \mathbb{C}^m \otimes \mathbb{C}^n$ by the identification (21) again. Then we have
$$\langle A, \phi_V \rangle = \sum |(z_i \,|\, V)|^2 = 0$$
by the relation (23), and $\phi_V \in A'$. Since $A' = \sigma(D, E)$, we have
$$\phi_V \in A' \cap \mathbb{P}_{m \wedge n} = \sigma(D, E) \cap \mathbb{P}_{m \wedge n} = \sigma_D.$$
This implies $V \in D$, and so we have $\mathcal{R}A = D^\perp$. For the second relation $\mathcal{R}A^\tau = E^\perp$, we note the following identities
$$\langle A^\tau, \phi_V \rangle = \langle A, \phi^V \rangle, \qquad \langle A^\tau, \phi^W \rangle = \langle A, \phi_W \rangle.$$
These imply that $A' = \sigma(D, E)$ if and only if $(A^\tau)' = \sigma(E, D)$. Therefore, the second relation $\mathcal{R}A^\tau = E^\perp$ follows from the first. $\square$

We will say that a pair $(D, E)$ is an *exposed decomposition pair* if it is a decomposition pair and $\sigma(D, E)$ is an exposed face.

**Theorem 6.2.** *Let $(D, E)$ be a pair of subspaces of $m \times n$ matrices. Then the following are equivalent:*

   (i) $(D, E)$ *is an exposed decomposition pair.*
   (ii) $(D^\perp, E^\perp)$ *is an intersection pair.*

*If this is the case then we have $\sigma(D, E) = \tau(D^\perp, E^\perp)'$.*

*Proof.* Suppose that the face $\sigma(D, E)$ is exposed, and take an element $A \in \text{int}\, \sigma(D, E)'$. Then we have
$$A' = \sigma(D, E)'' = \sigma(D, E)$$
by assumption. This implies that $\mathcal{R}A = D^\perp$ and $\mathcal{R}A^\tau = E^\perp$ by Lemma 6.1, and so we see that $A \in \text{int}\, \tau_{D^\perp} \cap \text{int}\, \tau^{E^\perp}$ by (24). This proves the relation
$$\text{int}\, (\tau_{D^\perp} \cap \tau^{E^\perp}) = \text{int}\, \sigma(D, E)' \subset \text{int}\, \tau_{D^\perp} \cap \text{int}\, \tau^{E^\perp}$$
by the relation (38). Therefore, we see that $(D^\perp, E^\perp)$ is an intersection pair.

For the converse, suppose that $(D^\perp, E^\perp)$ is an intersection pair. First of all, we see that $\sigma_D + \sigma^E = \tau(D^\perp, E^\perp)'$ is an exposed face of $\mathbb{D}$ by (38). We may take a decomposition pair $(D_1, E_1)$ such that $\sigma_D + \sigma^E = \sigma(D_1, E_1)$. It suffices to show that $D = D_1$ and $E = E_1$. To do this, take $A \in \text{int}\, \tau(D^\perp, E^\perp)$. Then we have $A \in \text{int}\, \tau_{D^\perp} \cap \text{int}\, \tau^{E^\perp}$ since $(D^\perp, E^\perp)$ is an intersection pair, and so
$$D^\perp = \mathcal{R}A, \qquad E^\perp = \mathcal{R}A^\tau,$$

by (24). On the other hand, we also have $A' = \tau(D^\perp, E^\perp)' = \sigma(D_1, E_1)$, and
$$D_1^\perp = \mathcal{R}A, \qquad E_1^\perp = \mathcal{R}A^\tau,$$
by Lemma 6.1, again. Therefore, we have $D = D_1$ and $E = E_1$. □

**Proposition 6.3.** *A pair $(D, E)$ of subspaces of $M_{m \times n}$ is an intersection pair if and only if there exists $A \in \mathbb{T}$ such that $\mathcal{R}A = D$ and $\mathcal{R}A^\tau = E$.*

*Proof.* Let $(D, E)$ be an intersection pair and take $A \in \operatorname{int} \tau(D, E)$. Then $A' = \tau(D, E)' = \sigma(D^\perp, E^\perp)$, and we have $\mathcal{R}A = D$ and $\mathcal{R}A^\tau = E$ by Lemma 6.1. For the converse, assume that there is $A \in \mathbb{T}$ such that $\mathcal{R}A = D$ and $\mathcal{R}A^\tau = E$. Take the intersection pair $(D_1, E_1)$ such that $A \in \operatorname{int} \tau(D_1, E_1)$. Then we have $\mathcal{R}A = D_1$ and $\mathcal{R}A^\tau = E_1$, and so $D = D_1$ and $E = E_1$. □

**Theorem 6.4.** *Every face of the convex cone $\mathbb{T}$ is exposed.*

*Proof.* Every face of $\mathbb{T}$ is of the form $\tau(D, E)$ for an intersection pair $(D, E)$ of spaces of matrices. Then $\sigma(D^\perp, E^\perp) = \tau(D, E)'$ by Theorem 6.2. Therefore, we have
$$\tau(D, E)'' = \sigma(D^\perp, E^\perp)' = \tau_D \cap \tau^E = \tau(D, E)$$
by (38). □

In the case of $m = n = 2$, every decomposition pair has been characterized in [16]. Since every positive map in $\mathbb{P}_1[M_2, M_2]$ is decomposable, this gives us the complete facial structures of the cone $\mathbb{P}_1[M_2, M_2]$. We just list up all of them:

| | | |
|---|---|---|
| I | $(3, 3)$ | $D = (xy^*)^\perp$, $E = (\bar{x}y^*)^\perp$ |
| II | $(2, 2)$ | $D = \{xy^*, zw^*\}^\perp$, $E = \{\bar{x}y^*, \bar{z}w^*\}^\perp$ ($x \nparallel z$ or $y \nparallel w$) |
| III | $(2, 2)$ | $D, E :$ has a unique rank one matrix |
| IV | $(2, 1)$ | $D$ has a unique rank one matrix, $E$ is spanned by a rank one matrix |
| V | $(1, 2)$ | $D$ is spanned by a rank one matrix, $E$ has a unique rank one matrix |
| VI | $(1, 1)$ | $D, E$ are spanned by rank two matrices |
| VII | $(1, 1)$ | $D = \mathbb{C}xy^*, E = \mathbb{C}\bar{x}y^*$ |
| VIII | $(1, 0)$ | $D$ is spanned by a rank two matrix, $E = \{0\}$ |
| IX | $(0, 1)$ | $D = \{0\}$, $E$ is spanned by a rank two matrix |

Here, the second column denotes the dimensions of $D$ and $E$, and $x \nparallel z$ means that $x$ is not parallel to $z$. We note that every 2-dimensional subspace of $M_{2 \times 2}$ has a rank one matrix. It is either spanned by rank one matrices, or it has a unique rank one matrix up to scalar multiplications. The space
$$D = \operatorname{span}\{e_{11} + e_{22}, e_{12}\}$$

is a typical example of the latter case. We remark that the faces of type I exhaust all maximal faces, and faces of type II (respectively VII) are the intersection of two (respectively three) maximal faces. The pairs

$$(D, D), \qquad (D, \mathbb{C}e_{12})$$

are typical examples of types III and IV, respectively. The faces of types III, IV and V are unexposed. Faces of types IV, VII and VIII (respectively V, VII and IX) consist of completely positive (respectively completely copositive) linear maps. The faces of types VII, VIII and IX are extreme rays of the cone $\mathbb{P}[M_2, M_2]$. Finally, faces of type II have different shapes according to whether $D$ consists of rank one matrices or not. Note that $D$ consists of rank one matrices if and only if $x \parallel z$ or $y \parallel w$. In this case, a face of type II is affine isomorphic to the cone $M_2^+$ of all positive semi-definite $2 \times 2$ matrices.

We note that Størmer [111] characterized in the sixties all extreme points of the convex set consisting of unital positive linear maps between $M_2$, whose facial structures can be found in [80].

The facial structures for the cone $\mathbb{V}_1(M_2 \otimes M_2)$ is now clear. One may take the dual faces from the above list except for unexposed cases. All possible nontrivial intersection pairs may be listed by the following table. When a given space is spanned by product vectors, we use 'SP' on the list. On the other hand, 'CE' means that the space is completely entangled.

|  |  | $D$ | $E$ | $D^\perp$ | $E^\perp$ |
|---|---|---|---|---|---|
| I' | $(1,1)$ | SP | SP | SP | SP |
| II' | $(2,2)$ | SP | SP | SP | SP |
| VI' | $(3,3)$ | SP | SP | CE | CE |
| VII' | $(3,3)$ | SP | SP | SP | SP |
| VIII' | $(3,4)$ | SP | SP | CE | $\{0\}$ |
| IX' | $(4,3)$ | SP | SP | $\{0\}$ | CE |

We have two cases for the type II'. Suppose that the pair of spaces are spanned by

$$x \otimes y, \ z \otimes w \qquad \text{and} \qquad \bar{x} \otimes y, \ \bar{z} \otimes w,$$

respectively. If $x \nparallel z$ and $y \nparallel w$ then we see that $x \otimes y$ and $z \otimes w$ are only product vectors in the span of them. Therefore, the corresponding face is the convex hull of two extreme rays. If we normalize them and consider the convex set of all separable states, then the resulting face is a one dimensional simplex. For faces of separable states which are simplices in higher dimensional cases, see [1] and [52]. If $x \parallel z$ and $y \parallel w$ then the resulting face is not a simplex.

It is not so easy in general to determine if a given pair of subspaces gives rise to a face of the cone $\mathbb{D}$ or not. This question has a close relation with the notion of optimality of entanglement witnesses, as we will see in Theorem 8.6. We close this section to characterize faces $\sigma(D, E)$ which are exposed by elements of the cone $\mathbb{V}_1$. Let
$$A = z_1 z_1^* + z_2 z_2^* \cdots + z_\iota z_\iota^* \in \mathbb{V}_1$$
be given with product vectors $z_i = \xi_i \otimes \eta_i \in \mathbb{C}^m \otimes \mathbb{C}^n$ for $i = 1, 2, \ldots \iota$. Then for $V \in M_{m \times n}$ we see that the following relations

(39)
$$\begin{aligned} \langle A, \phi_V \rangle = 0 &\iff \xi_i \otimes \eta_i \perp V \text{ for each } i = 1, 2, \ldots, \iota, \\ \langle A, \phi^W \rangle = 0 &\iff \langle A^\tau, \phi_W \rangle = 0 \iff \bar{\xi}_i \otimes \eta_i \perp W \text{ for each } i = 1, 2, \ldots, \iota \end{aligned}$$

hold. First, suppose that $\sigma(D, E)$ is exposed by $A \in \mathbb{V}_1$, and so

(40)
$$A' = \sigma(D, E), \qquad A' \cap \mathbb{P}_{m \wedge n} = \sigma_D, \qquad A' \cap \mathbb{P}^{m \wedge n} = \sigma^E.$$

From the condition $\sigma_D = A' \cap \mathbb{P}_{m \wedge n}$, we have $V \in D$ if and only if $\phi_V \in A'$ if and only if $V$ is orthogonal to $\xi_i \otimes \eta_i$ for each $i$. Similarly, we also have $W \in E$ if and only if $W$ is orthogonal to $\bar{\xi}_i \otimes \eta_i$ for each $i = 1, 2, \ldots, \iota$. Therefore, we see that the relations

(41)
$$D = \{\xi_1 \otimes \eta_1, \ldots, \xi_\iota \otimes \eta_\iota\}^\perp, \qquad E = \{\bar{\xi}_1 \otimes \eta_1, \ldots, \bar{\xi}_\iota \otimes \eta_\iota\}^\perp$$

hold. Conversely, suppose that the pair $(D, E)$ given by (41). Then we have $\phi_V \in A'$ if and only if $\langle A, \phi_V \rangle = 0$ if and only if $V \in D$ by the assumption and (39). This means $A' \cap \mathbb{P}_{m \wedge n} = \sigma_D$. Similarly, we also have $A' \cap \mathbb{P}^{m \wedge n} = \sigma^E$. Therefore, we see that the relation (40) holds.

**Theorem 6.5.** *For a pair $(D, E)$ of subspaces, the following are equivalent:*

(i) *$(D, E)$ is a decomposition pair and the face $\sigma(D, E)$ is exposed by elements of $\mathbb{V}_1$.*
(ii) *There exists a family $\{\xi_i \otimes \eta_i\}$ of product vectors in $\mathbb{C}^m \otimes \mathbb{C}^n$ with the relation (41).*

We say that a pair $(D, E)$ of subspaces of $\mathbb{C}^m \otimes \mathbb{C}^n$ is said to *satisfy the range criterion* if there exists a family $\{\xi_i \otimes \eta_i\}$ of product vectors in $\mathbb{C}^m \otimes \mathbb{C}^n$ such that
$$D = \operatorname{span}\{\xi_1 \otimes \eta_1, \ldots, \xi_\iota \otimes \eta_\iota\}, \qquad E = \operatorname{span}\{\bar{\xi}_1 \otimes \eta_1, \ldots, \bar{\xi}_\iota \otimes \eta_\iota\}.$$

It should be noted that the dimension gap between two spaces in the pair satisfying the range criterion may be quite big. For example, we put
$$x_\alpha = (1, \alpha)^{\mathrm{t}} \in \mathbb{C}^2, \qquad y_\alpha = (1, \bar{\alpha}, \ldots, \bar{\alpha}^{n-1})^{\mathrm{t}} \in \mathbb{C}^n$$

for $\alpha \in \mathbb{C}$, and consider the space
$$D = \operatorname{span}\left\{ x_\alpha y_\alpha^* = \begin{pmatrix} 1 & \alpha & \cdots & \alpha^{n-1} \\ \alpha & \alpha^2 & \cdots & \alpha^n \end{pmatrix} : \alpha \in \mathbb{C} \right\}$$

spanned by rank one matrices in $M_{2 \times n}$. It is easily seen that $D$ is an $(n+1)$-dimensional subspace with
$$D^\perp = \operatorname{span}\{e_{1,j+1} - e_{2,j} : j = 1, 2, \ldots, n-1\}.$$

We note that $D^\perp$ is completely entangled. On the other hand, the set
$$\left\{ \bar{x}_\alpha y_\alpha^* = \begin{pmatrix} 1 & \alpha & \cdots & \alpha^{n-1} \\ \bar{\alpha} & \bar{\alpha}\alpha & \cdots & \bar{\alpha}\alpha^{n-1} \end{pmatrix} : \alpha \in \mathbb{C} \right\}$$

generates the whole space $M_{2\times n}$. Indeed, the set
$$\{\bar x_\alpha y_\alpha^* : \alpha = 0, r_1, r_2, \ldots, r_{n-1}, ir_1, ir_2, \ldots, ir_n\}$$
is a basis of $M_{2\times n}$ whenever $r_1, r_2, \ldots, r_n$ are nonzero distinct real numbers. See [22]. This shows that the pair $(D^\perp, \{0\})$ is a decomposition pair and $\sigma(D^\perp, \{0\})$ is a face of $\mathbb{D}$ which is exposed by separable states. This means that the face $\sigma_{D^\perp} = \sigma(D^\perp, \{0\})$ of the face $\mathbb{P}_{m\wedge n}$ is still a face of the bigger cone $\mathbb{D}$. It is not known if this is a face of the cone $\mathbb{P}_1$. It was shown in [8] that if $D^\perp$ is a completely entangled subspace of $M_{2\times n}$ then the pair $(D, M_{2\times n})$ always satisfies the range criterion. This is not the case for $M_{3\times 3}$, since generic 4-dimensional subspaces of $M_{3\times 3}$ is entangled but the orthogonal complements have six rank one matrices up to scalar multiples.

Finding an exposed face $\sigma(D, E)$ which is not exposed by separable states has a close relation with the notion of edge PPTES, which will be the main topic of the next section.

## 7. Entangled edge states with positive partial tansposes

Suppose that $\tau(D, E)$ is a nontrivial face of the cone $\mathbb{T}$ generated by all PPT states. Since $\mathbb{V}_1$ is a convex subset of $\mathbb{T}$, we have the following three cases:

(i) $\operatorname{int} \tau(D, E) \cap \mathbb{V}_1 \neq \emptyset$,
(ii) $\operatorname{int} \tau(D, E) \cap \mathbb{V}_1 = \emptyset$, but $\tau(D, E) \cap \mathbb{V}_1 \neq \emptyset$,
(iii) $\tau(D, E) \cap \mathbb{V}_1 = \emptyset$.

We note that $\tau(D, E)$ has an element of $\mathbb{V}_1$ in its interior if and only if the dual face $\sigma(D^\perp, E^\perp)$ is exposed by an element of $\mathbb{V}_1$. Therefore, we see by Theorem 6.5 that the case (i) occurs if and only if the pair $(D, E)$ satisfies the range criterion. On the other hand, the case (iii) occurs if and only if there exists no product vector in $D$ whose partial conjugate lies in $E$. It is apparent that the case (iii) is the most important to understand the whole features of the convex cone $\mathbb{T}$.

We say that a PPTES $A_0$ is an *edge* if the smallest face determined by it satisfies the condition (iii). In other word, $A_0$ is an edge if and only if it is an interior point of a face $\tau(D, E)$ with no intersection with $\mathbb{V}_1$. Geometrically, this says that $A_0 \in \mathbb{T} \setminus \mathbb{V}_1$ is an edge if and only if any line segment from a separable state to $A_0$ cannot be extended within the cone $\mathbb{T}$. From this, it is evident that any PPTES is a convex sum of a separable state and an edge. We also see that $A_0$ is an edge if and only if for any $\epsilon > 0$ and $A \in \mathbb{V}_1$ we have $A_0 - \epsilon A \notin \mathbb{T}$, as it was originally introduced by Lewenstein, Kraus, Cirac and Horodecki [88]. It is also clear that $A \in \mathbb{T}$ is an edge if and only if there does not exist a nonzero product vector $\xi \otimes \eta \in \mathcal{R}A$ such that $\bar\xi \otimes \eta \in \mathcal{R}A^\tau$. Therefore, an edge state is a PPT state which violates the range criterion in an extreme way.

We say that an edge $A$ is *of type* $(p, q)$ if the range dimension of $A$ is $p$ and the range dimension of $A^\tau$ is $q$. The entanglement (22) given by Choi [27] is a $3 \otimes 3$ edge of type $(4, 4)$. This section will be split into two subsections. In the first one, we exhibit two main methods to construct edges, one using unextendible product basis, and another one using the duality theory. In the second subsection, we classify edges by their types in low dimensional cases, and mention briefly on extreme PPT states.

7.1. **Construction of PPT entangled edge states.** In this subsection, we explain two methods to construct edges. One important method is to use the notion of *unextendible product basis*, which is an orthonormal set of product vectors in $\mathbb{C}^m \otimes \mathbb{C}^n$ whose orthogonal complement has no product vector. For a given unextendible product basis $\mathcal{U}$, consider the projection $P_\mathcal{U} \in M_m \otimes M_n$ onto the span of $\mathcal{U}$. Then it is clear that the set $\tilde{\mathcal{U}} = \{\bar{\xi} \otimes \eta : \xi \otimes \eta \in \mathcal{U}\}$ is also an unextentible product basis and $(P_\mathcal{U})^\tau = P_{\tilde{\mathcal{U}}}$ by the relation (17). Therefore, we see that the projection $I - P_\mathcal{U}$ onto the orthogonal complement of $\mathcal{U}$ is of PPT. Since there is no product vector in the range of $I - P_\mathcal{U}$, we have the following, as it was found by Bennett, DiVincenzo, Mor, Shor, Smolin and Terhal [13].

**Theorem 7.1.** *If $\mathcal{U}$ is an unextendible product basis then $I - P_\mathcal{U}$ is a PPT entangled edge state.*

To get an example of an unextendible product basis, we consider the fifth roots of unity in the complex plane to get five vectors in $\mathbb{C}^3$

$$\xi_k = \lambda \left( \cos \frac{2\pi k}{5}, \sin \frac{2\pi k}{5}, h \right), \qquad k = 1, 2, 3, 4, 5,$$

where $h = \frac{1}{2}\sqrt{1 + \sqrt{5}}$ is chosen so that adjacent vectors are orthogonal to each other, and $\lambda = \dfrac{2}{\sqrt{5 + \sqrt{5}}}$ is chosen so that they are normal. Then it is easy to see that

$$\xi_k \otimes \xi_{2k \mod 5}, \qquad k = 1, 2, 3, 4, 5$$

forms an unextendible product basis in $\mathbb{C}^3 \otimes \mathbb{C}^3$. Another example is given by

$$e_1 \otimes (e_1 - e_2), \quad e_3 \otimes (e_2 - e_3), \quad (e_1 - e_2) \otimes e_3, \quad (e_2 - e_3) \otimes e_1,$$
$$(e_1 + e_2 + e_3) \otimes (e_1 + e_2 + e_3).$$

Unextendible product bases in $\mathbb{C}^3 \otimes \mathbb{C}^3$ have been completely characterized in [37], where the above two examples play key roles.

It is easy to see that six product vectors in $\mathbb{C}^3 \otimes \mathbb{C}^3$ are never orthogonal to each others, and so an unextendible product basis in $\mathbb{C}^3 \otimes \mathbb{C}^3$ has at most five vectors. We recall that any 5-dimensional subspace of $\mathbb{C}^3 \otimes \mathbb{C}^3$ has a product vector, and so an unextendible product basis in $\mathbb{C}^3 \otimes \mathbb{C}^3$ has exactly five vectors. Therefore, any $3 \otimes 3$ edge given by Theorem 7.1 gives rise to an edge of type $(4, 4)$. Many efforts have been made to understand $3 \otimes 3$ edge states of rank four. See [17, 54, 85, 86, 110], for example. It was shown recently by Chen and Djoković [18], and Skowronek [107] independently that all $3 \otimes 3$ PPT entangled states of rank four arise essentially from unextendible product bases. More precisely, they showed that every $3 \otimes 3$ PPTES of rank four is of the form

$$(U \otimes V)(I - P_\mathcal{U})(U \otimes V)^*$$

for an unextendible product basis $\mathcal{U}$ and nonsingular $3 \times 3$ matrices $U$ and $V$. See also [20] for more recent progresses in this direction. Especially, every $3 \otimes 3$ PPTES of rank four is an edge of type $(4,4)$.

Another useful method to construct edges of other types is to use the duality between states and maps. We begin with the early example of PPT entangled state found by Størmer [112] which also turns out to be an edge which is of type $(7,6)$. He gave an example of $A \in \mathbb{T}$ in order to give a short proof that the map $\Phi[1,0,\lambda]$ is an indecomposable positive linear map for $\lambda \geq 1$. This is given by

$$(42) \quad A = \begin{pmatrix} 2\mu & \cdot & \cdot & \cdot & 2\mu & \cdot & \cdot & \cdot & 2\mu \\ \cdot & 4\mu^2 & \cdot & \cdot & \cdot & \cdot & \cdot & \cdot & \cdot \\ \cdot & \cdot & 1 & \cdot & \cdot & \cdot & \cdot & \cdot & \cdot \\ \cdot & \cdot & \cdot & 1 & \cdot & \cdot & \cdot & \cdot & \cdot \\ 2\mu & \cdot & \cdot & \cdot & 2\mu & \cdot & \cdot & \cdot & 2\mu \\ \cdot & \cdot & \cdot & \cdot & \cdot & 4\mu^2 & \cdot & \cdot & \cdot \\ \cdot & \cdot & \cdot & \cdot & \cdot & \cdot & 4\mu^2 & \cdot & \cdot \\ \cdot & \cdot & \cdot & \cdot & \cdot & \cdot & \cdot & 1 & \cdot \\ 2\mu & \cdot & \cdot & \cdot & 2\mu & \cdot & \cdot & \cdot & 2\mu \end{pmatrix}.$$

If we identify $\mathbb{C}^3 \otimes \mathbb{C}^3$ and $M_{3\times 3}$ in the usual way, then we see that

$$\mathcal{R}A = \{e_{11} - e_{22},\ e_{22} - e_{33}\}^{\perp},$$
$$\mathcal{R}A^{\tau} = \{e_{12} - 2\mu e_{21},\ e_{23} - 2\mu e_{32},\ e_{31} - 2\mu e_{13}\}^{\perp}.$$

By a direct calculation, we see that there exists no nonzero product vector $\xi \otimes \eta \in \mathcal{R}A$ such that $\bar\xi \otimes \eta \in \mathcal{R}A^{\tau}$, when $\mu \neq \frac{1}{2}$. First of all, we note that

$$(43) \qquad B \perp \xi\eta^* \iff B\eta \perp \xi$$

for $B$ and $\xi\eta^*$ in $M_{m \times n}$. By (43), we see that $\xi\eta^* \in \mathcal{R}A$ if and only if

$$\xi_1 \bar\eta_1 = \xi_2 \bar\eta_2 = \xi_3 \bar\eta_3,$$

and $\bar\xi\eta^* \in \mathcal{R}A^{\tau}$ if and only if

$$\xi_1 \eta_2 = 2\mu \xi_2 \eta_1, \quad \xi_2 \eta_3 = 2\mu \xi_3 \eta_2, \quad \xi_3 \eta_1 = 2\mu \xi_1 \eta_3.$$

From this, we see that $\xi_1 \xi_2 \xi_3 \eta_1 \eta_2 \eta_3 = 0$, and we conclude that there is no rank one matrix $\xi\eta^* \in \mathcal{R}A$ with $\bar\xi\eta^* \in \mathcal{R}A^{\tau}$.

Now, we explain how to construct an edge from a given indecomposable positive linear map, as was done in [53] and [46]. Let $\sigma(D,E)$ be a proper face of the cone $\mathbb{D}$. Then we have the following two cases:

$$\operatorname{int} \sigma(D,E) \subset \operatorname{int} \mathbb{P}_1 \quad \text{or} \quad \sigma(D,E) \subset \partial \mathbb{P}_1,$$

since $\sigma(D,E)$ is a convex subset of the cone $\mathbb{P}_1$.

**Theorem 7.2.** *Let $\sigma(D,E)$ be a proper face of the cone $\mathbb{D}$. Then we have*

$$(44) \qquad \operatorname{int} \sigma(D,E) \subset \operatorname{int} \mathbb{P}_1 \iff \sigma(D,E)' \cap \mathbb{V}_1 = \{0\}.$$

*Proof.* For the direction ($\Longrightarrow$), assume that $A \in \sigma(D, E)'$ and $A \neq 0$. Take an interior point $\phi$ of $\sigma(D, E)$. Then it is also an interior point of the cone $\mathbb{P}_1$. Therefore, there is $t > 1$ such that
$$\psi := (1-t)\mathrm{Tr} + t\phi \in \mathbb{P}_1.$$
Since Tr is an interior point of the cone $\mathbb{D}$ and $A \neq 0$, we have $\langle A, \mathrm{Tr} \rangle > 0$ by Proposition 5.1. Furthermore, we have $\langle A, \phi \rangle = 0$, since $A \in \sigma(D, E)'$. Therefore, we have
$$\langle A, \psi \rangle = (1-t)\langle A, \mathrm{Tr} \rangle + t\langle A, \phi \rangle = (1-t)\langle A, \mathrm{Tr} \rangle < 0.$$
This shows that $A \notin \mathbb{V}_1$ by the duality between $\mathbb{P}_1$ and $\mathbb{V}_1$.

For the reverse direction, it suffices to show that
$$\sigma(D, E) \subset \partial \mathbb{P}_1 \implies \sigma(D, E)' \cap \mathbb{V}_1 \supsetneq \{0\}.$$
To do this, suppose that $\sigma(D, E) \subset \partial \mathbb{P}_1$. Take $\phi \in \mathrm{int}\,\sigma(D, E)$, and take the face $F$ of $\mathbb{P}_1$ such that $\phi \in \mathrm{int}\, F$. We note that $F$ is a proper face of $\mathbb{P}_1$ since $\phi \in \partial \mathbb{P}_1$ by assumption. We also note that $F$ is a face of $\mathbb{P}_1 = (\mathbb{V}_1)^\circ$ and $\sigma(D, E)$ is a face of $\mathbb{D} = \mathbb{T}^\circ$, and so we have
$$F' = \{A \in \mathbb{V}_1 : \langle A, \phi \rangle = 0\}$$
$$\sigma(D, E)' = \{A \in \mathbb{T} : \langle A, \phi \rangle = 0\}.$$
This shows that $\sigma(D, E)' \cap \mathbb{V}_1 = F'$, which has a nonzero element since $F$ is a proper face of $\mathbb{P}_1$. $\square$

The right side of (44) says that any nonzero element of $\sigma(D, E)'$ is an edge. Therefore, we conclude the following:

- If $\sigma(D, E)$ is a face of $\mathbb{D}$ with $\mathrm{int}\,\sigma(D, E) \subset \mathrm{int}\,\mathbb{P}_1$ then every nonzero element in the dual face $\sigma(D, E)'$ gives rise to an edge.
- Every edge state arises in this way.

The second claim follows from the fact that every face of the cone $\mathbb{T}$ is exposed by Theorem 6.4. Note that it is also possible to construct indecomposable positive maps using PPTES. See [121].

We begin with the map $\Phi[a, b, c]$ defined by (4) to construct edges of various types, as was done in [47]. Possible candidates satisfying the condition $\mathrm{int}\,\sigma(D, E) \subset \mathrm{int}\,\mathbb{P}_1$ is the case
$$0 < a < 2, \qquad 4bc = (2-a)^2, \qquad b \neq c.$$
If we fix $b$ and $c$, then we see that the family $\{\Phi[a, b, c] : 0 \le a \le 2\}$ is a line segment, and so it suffices to consider the map $\Phi[1, b, c]$, with the condition
$$4bc = 1, \qquad b \neq c.$$
We see that these maps are indeed interior points of the $\mathbb{P}_1$. To see this, we fix an interior point $x_0$ of a convex set $\mathbb{P}_1$, say the trace map in the cone $\mathbb{P}_1$, and recall [77] that $x$ is an interior point of $\mathbb{P}_1$ if and only if the line segment from $x_0$ to $x$ can be extended inside of $\mathbb{P}_1$. With this characterization, we see that an interior point of the 3-dimensional body for $\mathbb{P}_1$ described in Theorem 1.2 is really an interior point of the cone $\mathbb{P}_1$. Note that

$$\Phi[1,b,c] = \frac{1}{2}\Phi[2,0,0] + \frac{1}{2}\Phi\left[0,\sqrt{\frac{b}{c}},\sqrt{\frac{c}{b}}\right]$$

$$= \phi_{e_{11}-e_{22}} + \phi_{e_{22}-e_{33}} + \phi_{e_{33}-e_{11}} + \phi^{\mu e_{12}-\lambda e_{21}} + \phi^{\mu e_{23}-\lambda e_{32}} + \phi^{\mu e_{31}-\lambda e_{13}},$$

with $\lambda = \left(\frac{b}{c}\right)^{1/4}$ and $\mu = \left(\frac{c}{b}\right)^{1/4}$, and so $\lambda\mu = 1$ and $\lambda \neq 1$. Put

$$D = \{e_{11} - e_{22},\ e_{22} - e_{33},\ e_{33} - e_{11}\}^{\perp},$$
$$E = \{\mu e_{12} - \lambda e_{21},\ \mu e_{23} - \lambda e_{32},\ \mu e_{31} - \lambda e_{13}\}^{\perp}.$$

Then we see that every element of the dual face $\{\Phi[1,b,c]\}' = \sigma(D, E)$ gives rise to an edge. We note that $D$ and $E$ are the 7 and 6-dimensional spaces given by

$$D = \operatorname{span}\{e_{11} + e_{22} + e_{33},\ e_{12},\ e_{21},\ e_{23},\ e_{32},\ e_{31},\ e_{13}\},$$
$$E = \operatorname{span}\{e_{11},\ e_{22},\ e_{33},\ \lambda e_{12} + \mu e_{21},\ \lambda e_{23} + \mu e_{32},\ \lambda e_{31} + \mu e_{13}\},$$

respectively.

Typical examples in $\{\Phi[1,b,c]\}' = \tau(D, E)$ are given by

$$(45) \qquad X = \begin{pmatrix} 1 & \cdot & \cdot & \cdot & 1 & \cdot & \cdot & \cdot & 1 \\ \cdot & \lambda^2 & \cdot & (\eta|\xi) & \cdot & \cdot & \cdot & \cdot & \cdot \\ \cdot & \cdot & \mu^2 & \cdot & \cdot & \cdot & (\zeta|\xi) & \cdot & \cdot \\ \cdot & (\xi|\eta) & \cdot & \mu^2 & \cdot & \cdot & \cdot & \cdot & \cdot \\ 1 & \cdot & \cdot & \cdot & 1 & \cdot & \cdot & \cdot & 1 \\ \cdot & \cdot & \cdot & \cdot & \cdot & \lambda^2 & \cdot & (\zeta|\eta) & \cdot \\ \cdot & \cdot & (\xi|\zeta) & \cdot & \cdot & \cdot & \lambda^2 & \cdot & \cdot \\ \cdot & \cdot & \cdot & \cdot & \cdot & (\eta|\zeta) & \cdot & \mu^2 & \cdot \\ 1 & \cdot & \cdot & \cdot & 1 & \cdot & \cdot & \cdot & 1 \end{pmatrix}$$

with arbitrary unit vectors $\xi, \eta, \zeta$. Note that the partial transpose is given by

$$X^{\tau} = \begin{pmatrix} 1 & \cdot & \cdot & \cdot & (\xi|\eta) & \cdot & \cdot & \cdot & (\xi|\zeta) \\ \cdot & \lambda^2 & \cdot & 1 & \cdot & \cdot & \cdot & \cdot & \cdot \\ \cdot & \cdot & \mu^2 & \cdot & \cdot & \cdot & 1 & \cdot & \cdot \\ \cdot & 1 & \cdot & \mu^2 & \cdot & \cdot & \cdot & \cdot & \cdot \\ (\eta|\xi) & \cdot & \cdot & \cdot & 1 & \cdot & \cdot & (\eta|\zeta) & \cdot \\ \cdot & \cdot & \cdot & \cdot & \cdot & \lambda^2 & \cdot & 1 & \cdot \\ \cdot & \cdot & 1 & \cdot & \cdot & \cdot & \lambda^2 & \cdot & \cdot \\ \cdot & \cdot & \cdot & \cdot & \cdot & 1 & \cdot & \mu^2 & \cdot \\ (\zeta|\xi) & \cdot & \cdot & \cdot & (\zeta|\eta) & \cdot & \cdot & \cdot & 1 \end{pmatrix}.$$

We note that the rank of $X$ is equal to

$$1 + \operatorname{rank}\begin{pmatrix} (\xi|\xi) & (\xi|\eta) \\ (\eta|\xi) & (\eta|\eta) \end{pmatrix} + \operatorname{rank}\begin{pmatrix} (\eta|\eta) & (\eta|\zeta) \\ (\zeta|\eta) & (\zeta|\zeta) \end{pmatrix} + \operatorname{rank}\begin{pmatrix} (\zeta|\zeta) & (\zeta|\xi) \\ (\xi|\zeta) & (\xi|\xi) \end{pmatrix}$$

and the rank of $X^\tau$ is equal to
$$3 + \operatorname{rank} \begin{pmatrix} (\xi|\xi) & (\xi|\eta) & (\xi|\zeta) \\ (\eta|\xi) & (\eta|\eta) & (\eta|\zeta) \\ (\zeta|\xi) & (\zeta|\eta) & (\zeta|\zeta) \end{pmatrix}.$$
Recall that the rank of the $n \times n$ matrix $[(\xi_i|\xi_j)]_{i,j=1}^n$ is the dimension of the space span $\{\xi_1, \ldots, \xi_n\}$. We get an edge of

- type (7,6) if we take mutually independent vectors $\xi, \eta, \zeta$,
- type (7,5) if we take vectors so that $\dim \operatorname{span} \{\xi, \eta, \zeta\} = 2$ and none of two vectors are linearly dependent,
- type (6,5) if we take vectors so that $\dim \operatorname{span} \{\xi, \eta, \zeta\} = 2$ and one pair of two vectors are linearly dependent,
- type (4,4) if we take vectors with $\xi = \eta = \zeta$.

Note that the edge of type (4,4) obtained in this way with $\lambda = \sqrt{2}$ is nothing but the Choi's example (22). On the other hand, we get the Størmer's example (42) if we take orthonormal vectors $\{\xi, \eta, \zeta\}$. We may also get edges of type $(5,8)$ with variants of these examples.

7.2. **Classification of edges by their types.** It is easy to see that $A \in \mathbb{T}[M_m \otimes M_n]$ is of rank one then $A \in \mathbb{V}_1$. We proceed to find maximum rank of $A \in \mathbb{T}[M_m \otimes M_n]$ for which PPT implies automatically separability. Let $V$ and $W$ be subspaces of $\mathbb{C}^m$ and $\mathbb{C}^n$, respectively. We say that a positive semi-definite block matrix $A \in M_m \otimes M_n$ is *supported* on $V \otimes W$ if the range space of $A$ is contained in $V \otimes W$ and there is no proper subspace $V_0$ of $V$ or $W_0$ of $W$ such that the range is contained in $V_0 \otimes W_0$. If $A$ is supported on $V \otimes W$ which is a proper subspace of $\mathbb{C}^m \otimes \mathbb{C}^n$ then we can reduce the Hilbert spaces on which $A$ acts. For $A_1 \otimes A_2 \in M_m \otimes M_n$, we define the *partial traces* $\operatorname{Tr}_1$ and $\operatorname{Tr}_2$ by
$$\operatorname{Tr}_1(A_1 \otimes A_2) = \operatorname{Tr}(A_1)A_2, \qquad \operatorname{Tr}_2(A_1 \otimes A_2) = \operatorname{Tr}(A_2)A_1.$$
If $A = \sum_{i,j=1}^m e_{ij} \otimes A_{ij} \in M_m(M_n)$ then we have
$$\operatorname{Tr}_1 A = \sum_{k=1}^m A_{kk} \in M_n, \qquad \operatorname{Tr}_2 A = \sum_{i,j=1}^m (\operatorname{Tr} A_{ij}) e_{ij} \in M_m.$$
Since $(A_{kk} y | y) = (A(e_k \otimes y) | e_k \otimes y)$ for each $y \in \mathbb{C}^n$, we see that
$$y \in \operatorname{Ker}(\operatorname{Tr}_1 A) \iff x \otimes y \in \operatorname{Ker} A \text{ for each } x \in \mathbb{C}^m.$$
From this, we see that a positive semi-definite $A \in M_m \otimes M_n$ is supported on $V \otimes W$ if and only if both $\mathcal{R}(\operatorname{Tr}_1 A) = W$ and $\mathcal{R}(\operatorname{Tr}_2 A) = V$ hold. The ranks of $\operatorname{Tr}_1 A$ and $\operatorname{Tr}_2 A$ are called the *local ranks* of $A$. It should be noted that the local ranks of $A$ may be greater than the rank of $A$ itself, as we can see in the example (2). Nevertheless, it is known [59] that the rank of a PPT state is not less than the maximum of local ranks. Therefore, if a PPT state in $M_m \otimes M_n$ is supported on $\mathbb{C}^m \otimes \mathbb{C}^n$ then its rank is greater than or equal to $\max\{m, n\}$.

We consider the case of $m = 2$, and proceed to show that if $A$ is a $2 \otimes n$ PPT state of rank $n$ supported on $\mathbb{C}^2 \otimes \mathbb{C}^n$ then there exists a product vector $\xi \otimes \eta$ in the range of $A$ whose partial conjugate $\bar{\xi} \otimes \eta$ lies in the range of $A^\tau$. First of all, we know that every $n$-dimensional subspace $E$ of $M_{2 \times n}$ has at least one rank one matrix, as it was mentioned at the end of Section 3. This can be seen easily directly. To see this, take a basis $\{C_i\}$ of $E^\perp$. We are looking for $x^*y \in M_{2 \times n}$ which is orthogonal to each $C_i$, whose rows will be denoted by $C_i^1$ and $C_i^2$. If we write $x = (\alpha, \beta)$ then the orthogonality gives us the equation

$$0 = (C_i \,|\, x^*y) = (C_i^1 \,|\, \bar{\alpha}y) + (C_i^2 \,|\, \bar{\beta}y) = (\alpha C_i^1 + \beta C_i^2 \,|\, y)$$

for each $i = 1, 2, \ldots, n$. We denote by $C_{\alpha,\beta}$ the $n \times n$ matrix whose $i$th row is $\alpha C_i^1 + \beta C_i^2$. Then we can take $x = (\alpha, \beta)^t$ such that $C_{\alpha,\beta}$ is singular, and take $y$ such that $C_{\alpha,\beta} y = 0$.

Since $(x^t \xi_1 \,|\, \xi_2) = (x \bar{\xi}_2 \,|\, \bar{\xi}_1)$, we have the relation

$$(A^\tau(\xi_1 \otimes \eta_1) \,|\, \xi_2 \otimes \eta_2) = (A(\bar{\xi}_2 \otimes \eta_1) \,|\, \bar{\xi}_1 \otimes \eta_2),$$

in general. Especially, we see that $\xi \otimes \eta \in \operatorname{Ker} A$ if and only if $\bar{\xi} \otimes \eta \in \operatorname{Ker} A^\tau$ for a PPT state $A$. Now, we fix a product vector $\xi \otimes \eta \in \operatorname{Ker} A$ and take a unit vector $\xi_0 \in \mathbb{C}^2$ which is orthogonal to $\xi$. Then we see that $A(\xi_0 \otimes \eta) \neq 0$ by the assumption on the support. Furthermore, we have

$$(A(\xi_0 \otimes \eta) \,|\, \xi \otimes \omega) = (A^\tau(\bar{\xi} \otimes \eta) \,|\, \bar{\xi}_0 \otimes \omega) = 0$$

for every $\omega \in \mathbb{C}^n$. Therefore, we conclude that $A(\xi_0 \otimes \eta) = \xi_0 \otimes \zeta_1$ for a vector $\zeta_1 \in \mathbb{C}^n$. Similarly, we have $A^\tau(\bar{\xi}_0 \otimes \eta) = \bar{\xi}_0 \otimes \zeta_2$ for $\zeta_2 \in \mathbb{C}^n$. Now, for orthonomal basis $\{e_i\}$ of $\mathbb{C}^n$, we have

$$(\zeta_1 \,|\, e_i) = (\xi_0 \otimes \zeta_1 \,|\, \xi_0 \otimes e_i) = (A(\xi_0 \otimes \eta) \,|\, \xi_0 \otimes e_i)$$
$$= (A^\tau(\bar{\xi}_0 \otimes \eta) \,|\, \bar{\xi}_0 \otimes e_1) = (\bar{\xi}_0 \otimes \zeta_2 \,|\, \bar{\xi}_0 \otimes e_i) = (\zeta_2, e_i)$$

for each $i = 1, 2, \ldots, n$. This shows that $\zeta_1 = \zeta_2$, and we have the following:

**Proposition 7.3.** *Let $A \in M_2 \otimes M_n$ be a PPT state of rank $n$ supported on $\mathbb{C}^2 \otimes \mathbb{C}^n$. Then there exists $\xi_0 \in \mathbb{C}^2$ and $\eta, \zeta \in \mathbb{C}^n$ such that $A(\xi_0 \otimes \eta) = \xi_0 \otimes \zeta$ and $A^\tau(\bar{\xi}_0 \otimes \eta) = \bar{\xi}_0 \otimes \zeta$.*

This shows that if there is an edge of type $(p, q)$ supported on $\mathbb{C}^2 \otimes \mathbb{C}^n$ then $p, q > n$, to get a low bound for $p$ and $q$. The above proposition is one of the key arguments by Kraus, Cirac, Karnas and Lewenstein [73] who showed that any PPT states of rank $n$ supported on $\mathbb{C}^2 \otimes \mathbb{C}^n$ must be separable. More generally, it was shown in [58] that any PPT states of rank $\max\{m, n\}$ supported on $\mathbb{C}^m \otimes \mathbb{C}^n$ must be separable. From this, we have a lower bound for $p$ and $q$ for the existence of edges of type $(p, q)$.

**Theorem 7.4.** *Suppose that there is an edge of type $(p, q)$ supported on $\mathbb{C}^m \otimes \mathbb{C}^n$ then we have*

$$p, q > \max\{m, n\}.$$

In order to find upper bounds, we consider the following condition for a quadruplet $(k, \ell, m, n)$ of natural numbers:

(C) For any pair $(D, E)$ of subspaces of $\mathbb{C}^m \otimes \mathbb{C}^n$ with $\dim D^\perp = k$, $\dim E^\perp = \ell$, there exists a nonzero product vector $\xi \otimes \eta \in D$ with $\bar{\xi} \otimes \eta \in E$.

If the condition (C) holds then there is no edge of type $(mn - k, mn - \ell)$, which gives us upper bounds for range dimensions of an edge $A$ and its partial transpose $A^\tau$. We have the following [69]:

**Proposition 7.5.** *Let $(k, \ell, m, n)$ be a quadruplet of natural numbers with $k, \ell \leq mn$. If*

(46) $$(-\alpha + \beta)^k (\alpha + \beta)^l \neq 0 \quad \text{modulo} \quad \alpha^m, \beta^n,$$

*in the polynomial ring $\mathbb{Z}[\alpha, \beta]$, then the condition (C) holds.*

Precisely speaking, (46) means that $(-\alpha + \beta)^k (\alpha + \beta)^l$ is not contained in the ideal generated by $\alpha^m$ and $\beta^n$. The proof is an application of the intersection theory from algebraic geometry. If $k + \ell < m + n - 2$ then it can be shown that the condition (46) holds, and so the condition (C) always holds. Therefore, if $(mn - p) + (mn - q) < m + n - 2$ then there is no edge state of type $(p, q)$, in other word, if there is an edge state of type $(p, q)$ then we have
$$p + q \leq 2mn - m - n + 2.$$
In the case of $k + \ell = m + n - 2$, if
$$\sum_{r+s=m-1} (-1)^r \binom{k}{r}\binom{\ell}{s} \neq 0$$
then the condition (C) holds. Note that the left side is the coefficient of $\alpha^{m-1}\beta^{n-1}$ when we expand the polynomial (46). In this case, there is no edge state of type $(mn - k, mn - \ell)$. Note that the cases $k + \ell = m + n - 2$ are exactly the green lines of the figures in [85]. If $k + \ell > m + n - 2$ then the condition (C) does not hold, but this gives us no direct information for the existence of edge states. The Diophantine equation

(47) $$k + \ell = m + n - 2, \quad \sum_{r+s=m-1} (-1)^r \binom{k}{r}\binom{\ell}{s} = 0$$

is known as the Krawtchouk polynomial, which plays an important role in the coding theory. It is not yet solved completely. See [89] and [124]. We summarize as follows:

**Theorem 7.6.** *Suppose that there is an $m \otimes n$ edge of type $(p, q)$. Then we have the following:*

(i) $p + q \leq 2mn - m - n + 2$.
(ii) *If $p + q = 2mn - m - n + 2$ then $(k, \ell) = (mn - p, mn - q)$ satisfies the equation (47).*

We apply the above results to the $3 \otimes 3$ case. In this case, $2mn - m - n + 2 = 14$. When $k + \ell = 4$, the relation
$$\sum_{r+s=3-1} (-1)^r \binom{k}{r}\binom{\ell}{s} = 0$$

holds if and only if $(k, \ell) = (1, 3)$. Furthermore, every PPT entangled state of rank four must be an edge of type $(4, 4)$, as it was mentioned after Theorem 7.1. Therefore, we see that all possible types are

$$(4,4), \quad (5,5), \quad (5,6), \quad (5,7), \quad (6,6), \quad (5,8), \quad (6,7), \quad (6,8),$$

here we list up the cases $s \leq t$ by the symmetry. Edges of types $(5,5)$ and $(6,6)$ were found in [29] and [42] independently. Examples of edges of type $(6,8)$ have been constructed recently in [83], where all possible types of $3 \otimes 3$ edges were also constructed in a systematic way, except for $(4,4)$. This completes the classification of $3 \otimes 3$ edges by their types.

Now, we turn our attention to the $2 \otimes 4$ case. In this case, $2mn - m - n + 2 = 12$. When $k + \ell = 4$, we have

$$\sum_{r+s=2-1} (-1)^r \binom{k}{r} \binom{\ell}{s} = 0$$

if and only if $(k, \ell) = (2, 2)$. The case $(k, \ell) = (3, 1)$ is not a root of the equation, and this means that there is no edge of type $(5, 7)$. This special case was shown in [101]. Actually, all possible types are

$$(5,5), \quad (5,6), \quad (6,5), \quad (6,6).$$

The first example of PPTES given by Woronowicz [127] turns out to be an edge of type $(5, 5)$ in the $2 \otimes 4$ system. This example has been modified in [57] to get parameterized examples of the same type. Examples of edges of type $(5, 6)$ were found in [6]. It is still unknown whether there exists an edge of type $(6, 6)$ or not. We summarize in Figure 3.

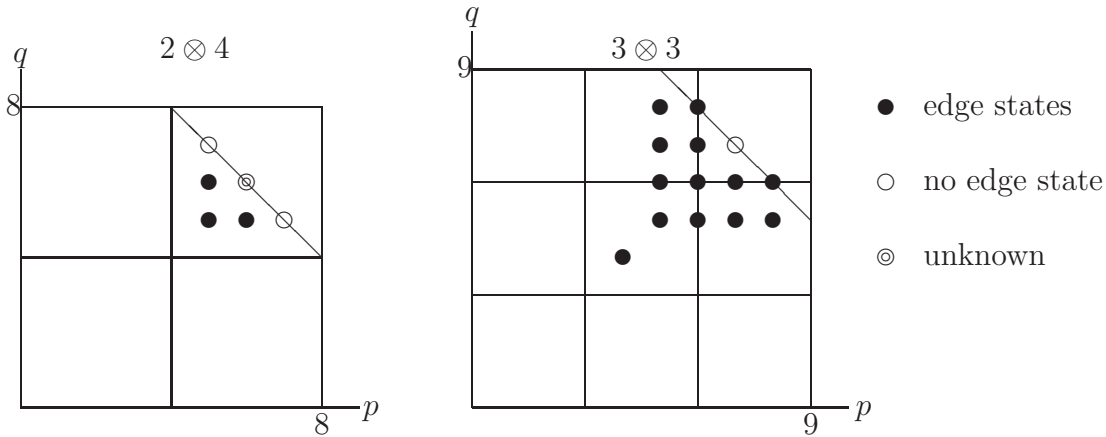

FIGURE 3. Possible types for $2 \otimes 4$ and $3 \otimes 3$ edges.

We close this section by mentioning briefly on extreme PPT states. A PPT state is said to be *extreme* if it generates an extreme ray of the cone $\mathbb{T}$. For a separable $A$, it is clear that $A$ is extreme if and only if it is of rank one. Since every face of the cone $\mathbb{T}$ is determined by a pair of subspaces, it is also apparent that every $3 \otimes 3$ edge of rank four is extreme. This is also the case for $2 \otimes 4$ edges of type $(5, 5)$. In the case of $2 \otimes 4$, it was shown in [6] that there is no extreme edge state of type (6,6). In the $3 \otimes 3$ case, edges of type $(5,5)$ and $(6,6)$ [29, 42] mentioned above were shown [43, 71] to be extreme.

An efficient method has been found [84] to check if a given face $\tau(D, E)$ is an extreme ray or not, where $D$ and $E$ are subspaces of $\mathbb{C}^m \otimes \mathbb{C}^n$. See also [44]. To explain this method, we consider the real Hilbert space $(M_m \otimes M_n)_h$ consisting of all $mn \times mn$ Hermitian matrices in $M_m \otimes M_n$ with the inner product $(X, Y) = \operatorname{Tr}(YX^{\mathrm{t}})$, and projections $P_D$ and $P_E$ in $(M_m \otimes M_n)_h$ onto $D$ and $E$, respectively. Define real linear maps $\phi_D$ and $\phi_E$ between $(M_m \otimes M_n)_h$ by

$$\phi_D(X) = P_D X P_D - X, \qquad \phi_B(X) = (P_E X^\tau P_E)^\tau - X, \qquad X \in (M_m \otimes M_n)_h.$$

Then we see that $\tau(D, E) \subset \operatorname{Ker} \phi_D \cap \operatorname{Ker} \phi_E$. Therefore, if $\operatorname{Ker} \phi_D \cap \operatorname{Ker} \phi_E$ is one-dimensional then $\tau(D, E)$ must be an extreme ray. It is not so difficult to see that this is also necessary for the extremeness of $\tau(D, E)$, to conclude that $\tau(D, E)$ is an extreme ray if and only if the relation

(48) $$\dim(\operatorname{Ker} \phi_D \cap \operatorname{Ker} \phi_E) = 1$$

holds. We note that the real dimension of $(M_m \otimes M_n)_h$ is just $(mn)^2$. We also note that $\operatorname{Ker} \phi_D$ consists of Hermitian matrices whose range is contained in $D$, and so the real dimension of $\operatorname{Ker} \phi_D$ is $(\dim D)^2$. Therefore, if $A$ is an $m \otimes n$ extreme PPT state of type $(p, q)$ then we have the inequality

$$p^2 + q^2 \leq (mn)^2 + 1,$$

by (48). In the case of $3 \times 3$, all possible types for extreme edges are given by

$$(4,4), \quad (5,5), \quad (5,6), \quad (5,7), \quad (6,6).$$

In a very recent paper [20], the authors checked extremeness for known examples to conclude that there are both extreme and non-extreme edges for types $(5,6)$, $(5,7)$ and $(6,6)$. It is not known if every $3 \otimes 3$ edge of type $(5,5)$ is extreme or not. For more systematic approach for extreme edges in higher dimensional cases, we refer to the recent paper [19].

## 8. Optimal entanglement witnesses

A Hermitian matrix $W$ is said to be an entanglement witness if there is entanglement $A_0$ with the property (18). Therefore, any entanglement witness is of the form $C_\phi^{\mathrm{t}}$ for a positive map $\phi$. Note that $C_\phi^{\mathrm{t}}$ is the Choi matrix $C_{\mathrm{tp}\circ\phi\circ\mathrm{tp}}$ of the map $\mathrm{tp} \circ \phi \circ \mathrm{tp}$, which is positive if and only if $\phi$ is positive. After Terhal [120] introduced the notion of entanglement witness, Lewenstein, Kraus, Cirac and Horodecki [87] studied the optimal entanglement witnesses which detect maximal sets of entanglement, and addressed [88] a fundamental question to find a minimal set of witnesses to detect all entanglement.

In this note, we say that a positive linear map $\phi$ *detects* entanglement $A$ if $\langle A, \phi \rangle < 0$, and $\phi$ is an *entanglement witness* if it detects entanglement. By duality, we see that a positive map $\phi$ is an entanglement witness if and only if it is not completely positive. We denote by $E_\phi$ the set of all entanglement detected by $\phi$, that is,
$$E_\phi := \{A \in (M_m \otimes M_n)^+ : \langle A, \phi \rangle < 0\}.$$
If $\lambda \phi_1 = \phi_2 + \psi$ for a $\lambda > 0$ and $\psi \in \mathbb{P}_{m \wedge n}$ then we have
$$\lambda \langle A, \phi_1 \rangle = \langle A, \phi_2 \rangle + \langle A, \psi \rangle, \qquad A \in (M_m \otimes M_n)^+.$$
Since $\langle A, \psi \rangle \geq 0$ for each $A \in (M_m \otimes M_n)^+$, we see that $\langle A, \phi_1 \rangle < 0$ implies $\langle A, \phi_2 \rangle < 0$, that is, $E_{\phi_1} \subset E_{\phi_2}$. Actually, the converse holds as was seen in [87]. The statement (iii) of the following proposition was pointed out by Kyung Hoon Han.

**Proposition 8.1.** *Let $\phi_1, \phi_2$ be entanglement witnesses. Then we have the following:*
  (i) $E_{\phi_1} \subset E_{\phi_2}$ *if and only if there is $\lambda > 0$ and $\psi \in \mathbb{P}_{m \wedge n}$ such that $\lambda \phi_1 = \phi_2 + \psi$.*
  (ii) $E_{\phi_1} \subsetneq E_{\phi_2}$ *if and only if there is $\lambda > 0$ and nonzero $\psi \in \mathbb{P}_{m \wedge n}$ such that $\lambda \phi_1 = \phi_2 + \psi$.*
  (iii) $E_{\phi_1} = E_{\phi_2}$ *if and only if there is $\lambda > 0$ such that $\lambda \phi_1 = \phi_2$.*

*Proof.* It suffices to show that $E_{\phi_1} \subset E_{\phi_2}$ implies that there exists $\lambda > 0$ such that
$$\lambda \langle A, \phi_1 \rangle \geq \langle A, \phi_2 \rangle, \qquad A \in (M_m \otimes M_n)^+, \tag{49}$$
since this would imply that $\lambda \phi_1 - \phi_2 \in \mathbb{P}_{m \wedge n}$ by the duality. First, we show the following
$$\langle A, \phi_1 \rangle = 0 \implies \langle A, \phi_2 \rangle \leq 0. \tag{50}$$
To see this, assume that $\langle A, \phi_1 \rangle = 0$ and $\langle A, \phi_2 \rangle > 0$ for $A \in (M_m \otimes M_n)^+$. If we take $B \in E_{\phi_1}$ then $\langle B + tA, \phi_1 \rangle < 0$ for any real $t$, but $\langle B + tA, \phi_2 \rangle \geq 0$ for sufficiently large $t$, contradictory to $E_{\phi_1} \subset E_{\phi_2}$. We note that
$$\langle A, \phi_1 \rangle < 0 \implies \left\langle A + \left|\frac{\langle A, \phi_1 \rangle}{\langle I_m \otimes I_n, \phi_1 \rangle}\right| I_m \otimes I_n, \phi_1 \right\rangle = 0$$
which implies
$$\left\langle A + \left|\frac{\langle A, \phi_1 \rangle}{\langle I_m \otimes I_n, \phi_1 \rangle}\right| I_m \otimes I_n, \phi_2 \right\rangle \leq 0$$
by (50). Therefore, we have
$$\langle A, \phi_1 \rangle < 0 \implies \frac{\langle I_m \otimes I_n, \phi_2 \rangle}{\langle I_m \otimes I_n, \phi_1 \rangle} \langle A, \phi_1 \rangle \geq \langle A, \phi_2 \rangle.$$
Finally, we consider the case when $\langle A, \phi_1 \rangle > 0$. We note that
$$\langle \langle A, \phi_1 \rangle B + |\langle B, \phi_1 \rangle| A, \phi_1 \rangle = 0$$
for every $B \in E_{\phi_1}$, which implies
$$\langle \langle A, \phi_1 \rangle B + |\langle B, \phi_1 \rangle| A, \phi_2 \rangle \leq 0,$$
by (50) again. From this, we get
$$\left|\frac{\langle B, \phi_2 \rangle}{\langle B, \phi_1 \rangle}\right| \geq \frac{\langle A, \phi_2 \rangle}{\langle A, \phi_1 \rangle}$$

for any $B \in E_{\phi_1}$. Therefore, we may put

$$\lambda = \inf \left\{ \left| \frac{\langle B, \phi_2 \rangle}{\langle B, \phi_1 \rangle} \right| : B \in E_{\phi_1} \right\}$$

to get (49).

Note that the statement (ii) is immediate from the statements (i) and (iii). For the statement (iii), suppose that $E_{\phi_1} = E_{\phi_2}$. Then there exist $\lambda_1, \lambda_2 > 0$ and $\psi_1, \psi_2 \in \mathbb{P}_{m \wedge n}$ such that

$$\lambda_1 \phi_1 = \phi_2 + \psi_1, \qquad \lambda_2 \phi_2 = \phi_1 + \psi_2,$$

which implies that

$$(\lambda_1 \lambda_2 - 1)\phi_1 = \lambda_2 \phi_2 + \lambda_2 \psi_1 - \phi_1 = \psi_2 + \lambda_2 \psi_1.$$

Since $\phi_1$ is not completely positive, we see that $\lambda_1 \lambda_2 - 1 = 0$ and $\psi_1 = \psi_2 = 0$. The converse is clear. □

An entanglement witness is said to be *optimal* if it detects a maximal set of entanglement. By Proposition 8.1, it is easy to describe the notion of optimality in terms of faces. We denote by $\mathbb{P}_\phi$ the smallest face of $\mathbb{P}_1$ containing $\phi$. Recall that this is the face in which $\phi$ is an interior point. The following was shown in [82], [103].

**Theorem 8.2.** *An entanglement witness $\phi \in \mathbb{P}_1$ is optimal if and only if there is no nonzero completely positive map in $\mathbb{P}_\phi$.*

*Proof.* If there is nonzero $\psi \in \mathbb{P}_{m \wedge n}$ such that $\psi \in \mathbb{P}_\phi$ then we see that $\phi = (1-t)\phi_2 + t\psi$ for $\phi_2 \in \mathbb{P}_\phi$ with $0 < t < 1$, since $\phi$ is an interior point of $\mathbb{P}_\phi$. This implies that $E_\phi \subsetneq E_{\phi_2}$, and so $\phi$ is not optimal. Conversely, if $\phi$ is not optimal then there is $\phi_2$ such that $E_\phi \subsetneq E_{\phi_2}$. Then there is $\lambda > 0$ and nonzero $\psi \in \mathbb{P}_{m \wedge n}$ such that $\lambda \phi = \phi_2 + \psi$. Since $\phi \in \mathbb{P}_\phi$ and $\mathbb{P}_\phi$ is a face we see that $\psi \in \mathbb{P}_\phi$. □

It is not so easy to determine if $\mathbb{P}_\phi$ has a completely positive map or not, since we do not know the facial structures of the cone $\mathbb{P}_1$ completely. But, it is easy to determine whether the bidual face $\{\phi\}''$, which is the smallest exposed face containing $\phi$, has a completely positive map or not. It should be noted that the dual is taken in the dual pair $(\mathbb{V}_1, \mathbb{P}_1)$. For example, $\{\phi\}'$ is a face of $\mathbb{V}_1$. We note that $\phi_V \in \{\phi\}''$ if and only if the following

(51) $$\langle zz^*, \phi \rangle = 0 \implies \langle zz^*, \phi_V \rangle = 0$$

holds. We define the set $P[\phi]$ of product vectors by

$$P[\phi] := \{z = \xi \otimes \eta \in \mathbb{C}^m \otimes \mathbb{C}^n : \langle zz^*, \phi \rangle = 0\}.$$

Then we see that

(52) $$\phi_V \in \{\phi\}'' \iff V \in P[\phi]^\perp$$

by the relation (23). This proves the equivalence between (i) and (iii) of the following:

**Proposition 8.3.** *Let $\phi \in \mathbb{P}_1$. Then the following are equivalent:*

(i) $\{\phi\}''$ has no nonzero completely positive map.
(ii) int $\{\phi\}' \subset$ int $\mathbb{V}_{m \wedge n}$.
(iii) The set $P[\phi]$ spans the whole space $\mathbb{C}^m \otimes \mathbb{C}^n$.

*Proof.* Note that $\{\phi\}' \subset \partial \mathbb{V}_{m \wedge n}$ if and only if there exists a nonzero $V$ such that $\{\phi\}' \subset \{\phi_V\}'$ since every convex set in the boundary lies in a maximal face. Note that the dual face $\{\phi_V\}'$ is taken with respect to the dual pair $(\mathbb{V}_{m \wedge n}, \mathbb{P}_{m \wedge n})$. Since the condition $\{\phi\}' \subset \{\phi_V\}'$ is also equivalent to (51), we have (i) $\iff$ (ii). □

We say that $\phi \in \mathbb{P}_1$ has the *spanning property* if it satisfies the conditions in Proposition 8.3. Therefore, if $\phi$ has the spanning property then $\phi$ is an optimal entanglement witness, as was seen in [87]. Note that the Choi map $\Phi[1, 0, 1]$ does not have the spanning property by (33), as was observed in [77]. See also [72]. Nevertheless, it is an optimal entanglement witness since it generates an extreme ray of the cone $\mathbb{P}_1$. Recently, it was shown in [7] that there exist examples of decomposable optimal entanglement witnesses without spanning properties.

For a product vector $z = \xi \otimes \eta$, we have
$$\langle zz^*, \phi^V \rangle = \langle (zz^*)^\tau, \phi_V \rangle = \langle (\bar\xi \otimes \eta)(\bar\xi \otimes \eta)^*, \phi_V \rangle.$$

Therefore, we see that $\{\phi\}''$ has no completely copositive map if and only if the partial conjugates of $P[\phi]$ span the whole space. If this is the case then we say that $\phi$ has the *co-spanning property*. It was shown in [22] that the Choi map $\Phi[1, 0, 1]$ has the co-spanning property. We also have the following:

**Proposition 8.4.** *Let $\phi \in \mathbb{P}_1$. Then the following are equivalent:*

(i) $\{\phi\}''$ has no nonzero completely copositive map.
(ii) int $\{\phi\}' \subset$ int $\mathbb{V}^{m \wedge n}$.
(iii) The partial conjugates of product vectors in $P[\phi]$ span the whole space $\mathbb{C}^m \otimes \mathbb{C}^n$.

We note that $\phi$ has the co-spanning property if and only if the composition $\phi \circ \mathrm{tp}$ with the transpose map has the spanning property. We also say that $\phi$ is *co-optimal* if $\phi \circ \mathrm{tp}$ is optimal. Very recently, the Choi type map (4) has been analyzed in [50] to find various examples which distinguish several notions of optimality. To do this, we first look at faces of the three dimensional convex body determined by the positivity condition given by Theorem 1.2 (i), as it is shown in Figure 4.

First of all, the convex body has four 2-dimensional faces: three of them are determined by $ab$, $bc$ and $ac$-planes; another one is determined by the plane $a + b + c = 2$. It is easy to see that they are neither optimal nor co-optimal by Theorem 1.2 (iii) and (iv). We also see five 1-dimensional faces: three of them come from $a$, $b$ and $c$-axes. They are neither optimal nor co-optimal. Another two 1-dimensional faces $e_{ab}$ and $e_{ac}$ are contained in the $ab$ and $ac$-plane, respectively. We also have a parameterized family $\{e_t\}$ of 1-dimensional faces which are the line segments between points on the circle parts and points on the hyperbola on the *bc*-plane. Now, 0-faces are clear. They consist of boundary points of Figure 2 and the hyperbola on the *bc*-plane. We summarize the results in Table 1.

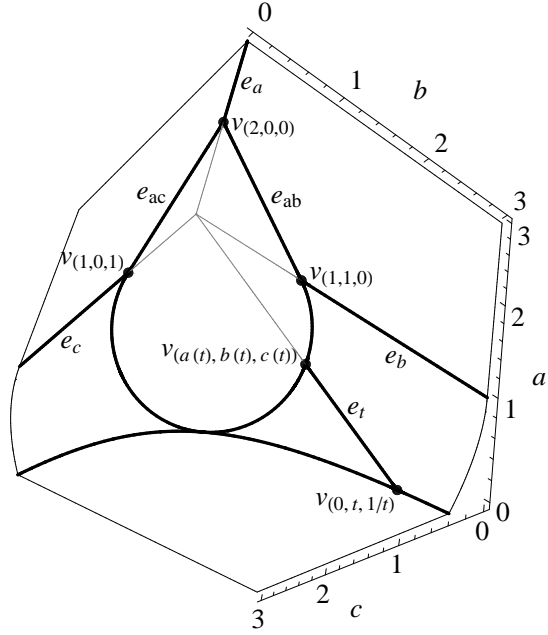

FIGURE 4. Part of convex body determined by Theorem 1.2 (i).

| Faces | Span. | Co-span. | Opt. | Co-opt. |
|---|---|---|---|---|
| $e_{ab}, e_{ac}, v_{(2,0,0)}$ | N | Y | N | Y |
| $e_t, v_{(0,t,1/t)}$ | Y | N | Y | N |
| $v_{(1,0,1)}, v_{(1,1,0)}$ | N | Y | Y | Y |
| $v_{(a(t),b(t),c(t))}$ | Y | Y | Y | Y |

TABLE 1. Summary of (co-)optimality and (co-)spanning property for faces of the convex body illustrated in Fig. 4.

Now, we turn our attention to optimal entanglement witnesses which detect PPTES. For a positive linear map $\phi \in \mathbb{P}_1$, we consider the set $E_\phi^\mathbb{T}$ of all PPT entanglement detected by $\phi$:
$$E_\phi^\mathbb{T} = \{A \in \mathbb{T} : \langle A, \phi \rangle < 0\}.$$
Note that $E_\phi^\mathbb{T}$ is nonempty if and only if $\phi$ is indecomposable by the duality between $\mathbb{T}$ and $\mathbb{D}$. The following theorem [49] tells us that exposed indecomposable positive maps detect quite large set of PPT entanglement with a nonempty interior. Recall that any entanglement is detected by an exposed positive linear map by Proposition 5.2. It should be noted that any dense subset of the set of all exposed positive maps also detects all entanglement.

**Theorem 8.5.** *For a positive linear map $\phi$, the following are equivalent:*

(i) $\phi$ has both the spanning and co-spanning properties.

(ii) $\{\phi\}''$ has no nonzero decomposable maps.

(iii) $\operatorname{int}\{\phi\}' \subset \operatorname{int} \mathbb{T}$.

(iv) The set $E_\phi^{\mathbb{T}}$ has the nonempty relative interior in $\mathbb{T}$.

(v) The set $E_\phi^{\mathbb{T}}$ contains a PPTES $A$ such that both $A$ and $A^\tau$ have the full ranges.

If $\phi$ is an exposed indecomposable positive linear map then the above conditions are automatically satisfied.

*Proof.* The implications (i) $\iff$ (ii) $\iff$ (iii) are consequences of Propositions 8.3 and 8.4. For (iii) $\implies$ (iv), take $A \in \operatorname{int}\{\phi\}'$ which is also an interior point of the cone $\mathbb{T}$. If we take a line segment from $I_m \otimes I_n$, which is an interior point of $\mathbb{T}$, to the boundary point $B$ of $\mathbb{T}$ through $A$, then any point $C$ on this line segment between $A$ and $B$ is an interior point of $\mathbb{T}$. It is now clear that $C$ is a relative interior point of $E_\phi^{\mathbb{T}}$ with respect to $\mathbb{T}$. The direction (iv) $\implies$ (v) is now clear.

It remains to prove the implication (v) $\implies$ (i). Suppose that both $A \in E_\phi^{\mathbb{T}}$ and $A^\tau$ have the full ranges, and consider the line segment between $A$ and the identity matrix $I_m \otimes I_n$. Since $\langle A, \phi \rangle < 0$ and $\langle I_m \otimes I_n, \phi \rangle > 0$, there is $A_0$ on the line segment such that $\langle A_0, \phi \rangle = 0$. Denote by $D$ and $E$ the orthogonal complements of the product vectors in $P[\phi]$ and the partial conjugates of product vectors in $P[\phi]$, respectively. Then $(D, E)$ is an exposed decomposition pair by Theorem 6.5, and so we see that $A_0$ belongs to the face $\sigma(D, E)' = \tau(D^\perp, E^\perp)$ of $\mathbb{T}$ by Theorem 6.2. Since $A_0$ is an interior point of $\mathbb{T}$, we conclude that both $D$ and $E$ are zeroes.

If $\phi$ is exposed then $\{\phi\}''$ is the ray generated by $\phi$. If $\phi$ is indecomposable then it is clear that $\{\phi\}''$ has neither completely positive nor completely copositive maps. This shows that $\phi$ satisfies both the spanning and co-spanning properties by Propositions 8.3 and 8.4. $\square$

It is now clear that decomposable exposed maps do not satisfy the conditions in Theorem 8.5. Even though decomposable maps cannot detect PPT entanglement, it is worthwhile to study those maps since they have a close relation to the facial structures of the cones $\mathbb{D}$ and $\mathbb{P}_1$. From now on, we suppose that $\phi$ is a decomposable map which is an optimal entanglement witness, and search conditions satisfied by $\phi$. To do this, we denote by $\mathbb{D}_\phi$ the smallest face of $\mathbb{D}$ containing $\phi$. First of all, the face $\mathbb{D} \cap \mathbb{P}_\phi$ of $\mathbb{D}$ has no completely positive maps, and so we see that $\mathbb{D} \cap \mathbb{P}_\phi = \sigma(0, E_1)$ for a subspace $E_1$. Since a completely copositive map $\phi^W$ is completely positive if and only if $W$ is of rank one, we see that $E_1$ must be completely entangled. In particular, $\phi$ must be completely copositive, and of the form

$$\phi = \phi^{W_1} + \phi^{W_2} + \cdots + \phi^{W_\nu}.$$

Since the relation $\mathbb{D}_\phi \subset \mathbb{D} \cap \mathbb{P}_\phi$ holds in general, $\mathbb{D}_\phi$ is of the form $\sigma(0, E_2)$ for a subspace $E_2$ of $E_1$, and so $\phi$ is an interior of $\sigma(0, E_2)$. We also note that the map $\phi$ is an interior point of the convex set $\sigma^{E_3}$ with $E_3 = \operatorname{span}\{W_1, \ldots, W_\nu\}$ by (26). Since $\sigma(0, E_2) = \sigma^{E_2}$ is a face of $\mathbb{P}^{m \wedge n}$, we conclude that $E_2 = E_3$, and
$$\mathbb{D}_\phi = \sigma(0, E), \qquad \text{where } E = \text{span}\{W_1, \ldots, W_\nu\}.$$

In this case, we say that $\phi$ is *supported* on the space $E = \text{span}\{W_1, W_2, \ldots, W_\nu\}$. In this way, we get the conditions (i) and (iii) in Theorem 8.6 below. We note that the condition (i) had been already known in [87]. To get another necessary condition, we note that if $\phi$ is optimal then it must be on the boundary of the cone $\mathbb{P}_1$, and so there exists a product vector $z = \xi \otimes \eta$ such that $\langle zz^*, \phi \rangle = 0$ by Corollary 5.6. Since

$$\langle zz^*, \phi^W \rangle = \langle (zz^*)^\tau, \phi_W \rangle = |(\bar\xi \otimes \eta | W)|^2$$

by (13) and (23), We have the relation $\sum_{i=1}^\nu |(\bar\xi \otimes \eta | W_i)|^2 = \langle zz^*, \phi \rangle = 0$. We summarize as in the following theorem [82].

**Theorem 8.6.** *Let $\phi$ be a completely copositive linear map supported on the subspace $E$ of $M_{m \times n} = \mathbb{C}^m \otimes \mathbb{C}^n$. If $\phi$ is an optimal entanglement witness then we have the following:*

(i) *$E$ is completely entangled.*
(ii) *$E^\perp$ has a product vector.*
(iii) *The convex set $\sigma^E$ is a face of $\mathbb{D}$.*

When $m = 2$, it was shown in [8] that a completely copositive map is an optimal entanglement witness if and only if it has the spanning property if and only if its support is completely entangled. Especially, we see that the condition (i) of Theorem 8.6 actually implies conditions (ii) and (iii).

In the case of $m = n = 3$, we can find an example of a completely copositive map supported on a completely entangled space which does not satisfy the condition (iii) of Theorem 8.6. See [82]. Recall that the dimension of completely entangled subspaces of $M_{m \times n}$ is at most $(m-1)(n-1)$. Therefore, if $m = 2$ or $m = n = 3$ then the condition (i) of Theorem 8.6 implies the condition (ii). In the case of $m = 3$ and $n = 4$, there are examples of 6-dimensional completely entangled subspace whose orthogonal complement is also completely entangled. See [7] and [108]. It would be interesting to determine if the converse of Theorem 8.6 holds or not.